\documentclass[11pt,a4paper]{article}
\pdfoutput=1
\usepackage{jheppub}

\usepackage{multirow, graphicx,amssymb,url,mathrsfs,amsmath}
\usepackage{wrapfig,boxedminipage,setspace,subfigure,epsfig}
\usepackage{amsxtra,amstext,latexsym,dsfont,amsfonts}
\usepackage{color,eucal}
\usepackage[dvipsnames]{xcolor}
\usepackage{float}
\usepackage{slashed,comment}
\usepackage{kotex}
\usepackage{tabularx, array}

\newcolumntype{L}[1]{>{\raggedright\arraybackslash}p{#1}}
\newcolumntype{C}[1]{>{\centering\arraybackslash}p{#1}}
\newcolumntype{R}[1]{>{\raggedleft\arraybackslash}p{#1}}

\newcommand{\dd}{\mathrm{d}}

\newcommand{\emu}{{\left(\frac{\partial{\epsilon}}{\partial{\mu}}\right)_{T}}}
\newcommand{\eT}{{\left(\frac{\partial{\epsilon}}{\partial{T}}\right)_{\mu}}}

\newcommand{\Pmu}{{\left(\frac{\partial{P}}{\partial{\mu}}\right)_{T}}}
\newcommand{\PT}{{\left(\frac{\partial{P}}{\partial{T}}\right)_{\mu} }}

\newcommand{\rhomu}{{\left(\frac{\partial{\rho}}{\partial{\mu}}\right)_{T}}}
\newcommand{\rhoT}{{\left(\frac{\partial{\rho}}{\partial{T}}\right)_{\mu}}}

\title{Quasi-normal modes of dyonic black holes and magneto-hydrodynamics}

\author[a,b]{Hyun-Sik Jeong,}
\author[c,d]{Keun-Young Kim,}
\author[a,b]{and Ya-Wen Sun}

\emailAdd{hyunsik@ucas.ac.cn}
\emailAdd{fortoe@gist.ac.kr}
\emailAdd{yawen.sun@ucas.ac.cn}

\affiliation[a]{School of physics $\&$ CAS Center for Excellence in Topological Quantum Computation, University of Chinese Academy of Sciences, Zhongguancun east road 80, Beijing 100049, China}
\affiliation[b]{Kavli Institute for Theoretical Sciences, University of Chinese Academy of Sciences, \\ Zhongguancun east road 80, Beijing 100049, China}
\affiliation[c]{Department of Physics and Photon Science, Gwangju Institute of Science and Technology, \\
123 Cheomdan-gwagiro, Gwangju 61005, Korea}
\affiliation[d]{Research Center for Photon Science Technology, Gwangju Institute of Science and Technology, \\
123 Cheomdan-gwagiro, Gwangju 61005, Korea}

\abstract{
We revisit the magneto-hydrodynamics in (2+1) dimensions and confirm that it is consistent with the quasi-normal modes of the (3+1) dimensional dyonic black holes in the most general set-up with finite density, magnetic field and wave vector. We investigate all possible modes (sound, shear, diffusion, cyclotron etc.) and their interplay. For the magneto-hydrodynamics we perform a complete and detailed analysis correcting some prefactors in the literature, which is important for the comparison with quasi-normal modes. For the quasi-normal mode computations in holography we identify the independent fluctuation variables of the dyonic black holes, which is nontrivial at finite density and magnetic field. As an application of the quasi-normal modes of the dyonic black holes we investigate a transport property, the diffusion constant. We find that the diffusion constant at finite density and magnetic field saturates the lower bound at low temperature. We show that this bound can be understood from the pole-skipping point.
}

\begin{document}
\maketitle

%%%%%%%%%%%%%%%%%%%%%%%%%%%%%%%%
%    Section: Introduction
%%%%%%%%%%%%%%%%%%%%%%%%%%%%%%%%
\section{Introduction}

Holography (gauge/gravity duality) has provided useful methods to study properties of strongly coupled systems~\cite{Hartnoll:2016apf,Zaanen:2015oix,Ammon:2015wua,Baggioli:2019rrs}. In particular, the holographic descriptions of the strongly correlated (2+1) dimensional collective dynamics have been implemented to shed light on long-standing condensed matter problems such as the quantum phase transition~\cite{Hartnoll:2007ih}, superfluidity~\cite{Basu:2008st,Herzog:2008he}, high-temperature superconductivity~\cite{Hartnoll:2008vx,Hartnoll:2008kx}.\footnote{For recent developments of the holographic study for the holographic superconductivity, see \cite{Jeong:2021wiu,Baggioli:2022aft}.}

One of the milestones for the strongly coupled (2+1) dimensional field theories in holography is that the quasi-normal modes of the (3+1) dimensional AdS black holes are consistent with the predictions of (2+1) dimensional hydrodynamics, for instance, the holographic model with the explicitly (or spontaneously) broken translational symmetry~\cite{Davison:2014lua,Amoretti:2017frz,Andrade:2017cnc,Amoretti:2018tzw,Ammon:2019wci,Amoretti:2019cef,Amoretti:2019kuf,Baggioli:2021xuv,Jeong:2021zhz} and the superfluid where the U(1) symmetry is broken spontaneously~\cite{Amado:2009ts,Herzog:2009md,Yarom:2009uq,Herzog:2011ec,Amado:2013xya,Amado:2013aea,Esposito:2016ria,Arean:2021tks} or pseudo-spontaneously~\cite{Ammon:2021slb}.\footnote{See also \cite{Blake:2018leo,Arean:2020eus,Wu:2021mkk,Jeong:2021zsv,Jeong:2021zhz,Huh:2021ppg,Liu:2021qmt} for the study of the bound of diffusion constants from the linearized hydrodynamics using quasi-normal modes.}

Comparing the quasi-normal modes with hydrodynamic predictions will be an important and interesting research direction because it may not only provide more supporting (indirect) evidence of holographic duality, but also gives us novel analysis for the transport properties of strongly correlated systems. Note that hydrodynamics can tell which transport coefficients appear in the theory and holography reveals the details of the transport properties of such coefficients. 

In this paper, we study the (3+1) dimensional AdS black hole in the presence of external magnetic fields at finite density, \textit{dyonic} black holes (Einstein-Maxwell model), which is dual to the (2+1) dimensional quantum field theory in external magnetic fields.
In particular, we aim to compute the quasi-normal modes of dyonic black holes and compare them with the viscous magneto-hydrodynamics proposed by Hartnoll-Kovtun-M\"uller-Sachdev (HKMS)~\cite{Hartnoll:2007ih}. Thus, this paper is along the line of the developments of ``the comparison between the quasi-normal modes in (3+1) dimensions and the hydrodynamics in (2+1) dimensions" in holography~\cite{Davison:2014lua,Amoretti:2017frz,Andrade:2017cnc,Amoretti:2018tzw,Ammon:2019wci,Amoretti:2019cef,Amoretti:2019kuf,Baggioli:2021xuv,Jeong:2021zhz,Amado:2009ts,Herzog:2009md,Yarom:2009uq,Herzog:2011ec,Amado:2013xya,Amado:2013aea,Esposito:2016ria,Arean:2021tks,Ammon:2021slb}.

The dyonic black holes in (3+1) dimensions is one of well-studied black hole models in holography from the thermodynamic properties~\cite{Hartnoll:2007ih,Hartnoll:2007ip,Hartnoll:2007ai,Herzog:2007ij,Denef:2009yy} to the transport properties~\cite{OBannon:2007in,Buchbinder:2008dc,Buchbinder:2009aa,Buchbinder:2009mk,Bergman:2012na,Gubankova:2013lca,Dutta:2013dca,Kim:2015wba,Amoretti:2015gna,Blake:2015ina,Lucas:2015pxa,Zhou:2015dha,Davison:2015bea,Blake:2015hxa,Donos:2015bxe,Seo:2015pug,Ahn:2015shg,Amoretti:2016cad,Kim:2016hzi,Ge:2016sel,Khimphun:2017mqb,Blake:2017qgd,Cremonini:2017qwq,Seo:2017yux,Chen:2017gsl,Blauvelt:2017koq,Angelinos:2018qlc,Pal:2019bfw,Kim:2019lxb,Hoyos:2019pyz,Song:2019rnf,Amoretti:2019buu,Baggioli:2020edn,An:2020tkn,Kim:2020ozm22,Amoretti:2020mkp,Amoretti:2021lll,Amoretti:2021fch,Jokela:2021uws,Priyadarshinee:2021rch} such as the Hall conductivity, Nernst Effect, diverse magneto-transport and magnetic phase transition.\footnote{The magnetic susceptibility in holography turned out to be of order $1/T$ with temperature $T$, which is different from the weakly coupled systems such as the free electron gas: the magnetic susceptibility is independent of $T$. Thus, dyonic black holes in holography show the imprint of the strongly correlated field theories at finite magnetic fields.} 
However, surprisingly enough, a complete study of the quasi-normal mode excitations in dyonic black holes has been still lacking up to date.

In \cite{Jeong:2021zhz}, quasi-normal modes of magnetically charged black holes (i.e., \textit{zero} density) have been compared with the hydrodynamic theory only for the \textit{sound} channel.\footnote{For the quasi-normal modes of electrically charged black holes (zero magnetic fields), see \cite{Edalati:2010pn,Davison:2011uk,Gushterov:2018spg}.} As we will describe in the main context, at zero density, there will be two decoupled channels: the sound channel and the shear channel.
In this paper, we fill the gap for the complete study of quasi-normal modes of (3+1) dimensional dyonic black holes. In other words, we compute the quasi-normal modes from all the channels (sound channel, shear channel) at \textit{zero} density as well as the case at \textit{finite} density in which the sound channel is coupled with the shear channel and compare all quasi-normal modes with the HKMS magneto-hydrodynamics \cite{Hartnoll:2007ih}.\footnote{See also \cite{Baggioli:2021ujk} for the quasi-normal mode analysis at zero density in the presence of the strength of Coulomb interactions, \cite{Baggioli:2020edn,Donos:2021ueh,Amoretti:2021lll,Delacretaz:2019wzh} for the magneto-phonon in which the translational invariance is broken, and \cite{Brattan:2012nb,Janiszewski:2015ura,Ammon:2016fru,Ammon:2017ded,Ammon:2020rvg} for higher dimensional dyonic black holes.} 

%Note that when the system has both a finite density and a magnetic fields all the fluctuations (channels) are coupled each other. For such a case, the independent fluctuation variables associated with the diffeomorphism invariance together with gauge invariance has not been present so far. In this paper, we found such variables \eqref{GIVOUR} consistent with the one given in previous literature at \textit{zero} density (or \textit{zero} magnetic field) and we compute quasi-normal mode spectrum for the dyonic black holes.

In addition to checking quasi-normal modes of dyonic black holes with the HKMS magneto-hydrodynamics, we also study the transport property that appeared at finite wave vector: the diffusion constant. In particular, we focus on the bound of the diffusion constant of dyonic black holes and study its relation with the pole-skipping argument~\cite{Blake:2018leo,Jeong:2021zhz}.

This paper is organized as follows. 
In section \ref{SECMHDDIS}, we revisit the HKMS magneto-hydrodynamics in (2+1) dimensions in details. 
In section \ref{SECMHDDIS333}, we introduce (3+1) dimensional dyonic black holes as well as the method for quasi-normal modes computation: the determinant method. Then, implementing the determinant method, we compute the quasi-normal modes of dyonic black holes and compare them with the hydrodynamic predictions given in section \ref{SECMHDDIS}. Also we study the bound of the diffusion constant and the pole-skipping.
Section \ref{SECMHDDIS444} is devoted to conclusions.

%%%%%%%%%%%%%%%%%%%%%%%%%%%%%%%%%%%%%%
%    Section: 
%%%%%%%%%%%%%%%%%%%%%%%%%%%%%%%%%%%%%%
\section{Magneto-hydrodynamics revisited}\label{SECMHDDIS}

In this section, we revisit the viscous magneto-hydrodynamics in (2+1) dimensions in the presence of the density ($\rho$) and the magnetic field ($H$), which is proposed by Hartnoll-Kovtun-M\"uller-Sachdev (HKMS)~\cite{Hartnoll:2007ih}. 

It will be instructive to note that the main interest in \cite{Hartnoll:2007ih} is the transport properties at \textit{zero} wave vector such as conductivities~\cite{OBannon:2007in,Buchbinder:2008dc,Buchbinder:2009aa,Buchbinder:2009mk,Bergman:2012na,Kim:2015wba,Amoretti:2015gna,Lucas:2015pxa,Zhou:2015dha,Davison:2015bea,Blake:2015hxa,Donos:2015bxe,Seo:2015pug,Ahn:2015shg,Amoretti:2016cad,Kim:2016hzi,Ge:2016sel,Khimphun:2017mqb,Blake:2017qgd,Cremonini:2017qwq,Seo:2017yux,Chen:2017gsl,Blauvelt:2017koq,Angelinos:2018qlc,Pal:2019bfw,Kim:2019lxb,Hoyos:2019pyz,Song:2019rnf,Amoretti:2019buu,Baggioli:2020edn,An:2020tkn,Kim:2020ozm22,Amoretti:2020mkp,Amoretti:2021lll,Amoretti:2021fch,Priyadarshinee:2021rch}. In this paper, we study the properties of HKMS magneto-hydrodynamics at \textit{finite} wave vector. Thus we aim to study the complete analysis of the HKMS magneto-hydrodynamics. In particular, we focus on the dispersion relations as well as the transport properties that appeared at finite wave vector such as the diffusion constant.\footnote{Note that the magnetic field is assumed to be a fixed constant in the hydrodynamic limit in the HKMS \cite{Hartnoll:2007ih}. For interesting development for the case of vanishing magnetic fields in the hydrodynamic regime, see \cite{Buchbinder:2008dc,Buchbinder:2009aa,Buchbinder:2009mk}.} Note that, in the main context below, we will revise two things about the dispersion relations given in the appendix of \cite{Hartnoll:2007ih}: one is a sign typo and the other is the prefactor in the gapless hydrodynamic mode, which is important to be consistent with quasi-normal modes from holography.

\subsection{Equations of motion}

The equations of motion for hydrodynamics are the conservation laws:
\begin{align}\label{CONSERVATIONEQ}
\begin{split}
\partial^{\nu} T_{\mu\nu} = F_{\mu\nu} J^{\nu} \,, \qquad
\partial_{\mu} J^{\mu} = 0,
\end{split}
\end{align} 
where $F_{\mu\nu}$ is the field strength of the electromagnetic field, $T_{\mu\nu}$ is the stress tenser, and $J^{\mu}$ is the current.\footnote{See \cite{Hartnoll:2007ih} for the details of subtracting out the magnetization current.}
%and the coupling of the magnetization to the magnetic fields
In the case under consideration, we take $F_{\mu\nu}$ to be magnetic as
\begin{align}\label{}
\begin{split}
F_{tx} = 0\,, \quad F_{ty} = 0\,, \quad F_{ij} = \epsilon_{ij} H\,,
\end{split}
\end{align} 
where $i,j = (x, y)$.
%
%\paragraph{The stress tensor and current with the dissipative part:}
One can find the stress tensor $T_{\mu\nu}$ at first order in derivatives as
\begin{align}\label{TMUNUEQ}
\begin{split}
T^{\mu\nu}    \,=\, \epsilon u^{\mu} u^{\nu} + P \Delta^{\mu\nu} + \Pi^{\mu\nu} \,,
\end{split}
\end{align} 
where $\epsilon$ is the energy density, $P$ is the pressure, and $\Delta^{\mu\nu} = \eta^{\mu\nu} + u^{\mu} u^{\nu}$ with the fluid velocity $u^{\mu}$. $\Pi^{\mu\nu}$ is the dissipative term given by
\begin{align}\label{TMUNUEQ2}
\begin{split}
\Pi^{\mu\nu}  \,=\, - \eta\left[\Delta^{\mu\alpha} \Delta^{\nu\beta}(\partial_{\alpha} u_{\beta} + \partial_{\beta} u_{\alpha})  - \Delta^{\mu\nu} \partial_{\gamma}u^{\gamma} \right] \,,
\end{split}
\end{align} 
where $\eta$ is the shear viscosity.\footnote{There could be a bulk viscosity in the dissipative term, which is irrelevant for the conformally invariant theory considered in this paper. See details in \cite{Buchbinder:2008dc} for the unbroken conformal invariance in the presence of the gauge fields.} Note that $\Pi^{\mu\nu}$ is vanishing at local equilibrium by definition.

Similarly, the current $J^{\mu}$ can also be expressed at first order as 
\begin{align}\label{JCURRENTEQ}
\begin{split}
J^{\mu} = \rho u^{\mu} +  \nu^{\mu} \,,
\end{split}
\end{align} 
where $\rho$ is the charge density and $\nu^{\mu}$ is the dissipative part given by
\begin{align}\label{JCURRENTEQ2}
\begin{split}
\nu^{\mu} = \sigma \Delta^{\mu\nu} \left(-\partial_{\nu} \mu + F_{\nu\alpha} u^{\alpha} + \frac{\mu}{T}\partial_{\nu} T\right) \,,
\end{split}
\end{align} 
where $\sigma$ is the conductivity, $\mu$ is the chemical potential, and $T$ is the temperature.\footnote{\eqref{JCURRENTEQ2} can be obtained by the argument with the positive entropy production~\cite{Hartnoll:2007ih}.}

Choosing four independent variables $(\delta u_{i \,=\, x, y}, \, \delta T,\, \delta \mu)$, we study the fluctuations around the equilibrium in which 
\begin{align}\label{EQSE}
\begin{split}
u^{\mu} = (1, 0, 0) \,, \qquad T = \text{constant}\,, \qquad \mu = \text{constant} \,.
\end{split}
\end{align} 
Based on \eqref{EQSE}, one can find that the relevant fluctuations for \eqref{TMUNUEQ} and \eqref{JCURRENTEQ} are
\begin{align}\label{FLUCFLUC}
\begin{split}
\delta T^{tt} &= \left(\frac{\partial\epsilon}{\partial\mu}\right)_{T} \delta \mu + \left(\frac{\partial\epsilon}{\partial T}\right)_{\mu} \delta T \,, \qquad\qquad\qquad\qquad\qquad \delta T^{ti} = \left(\epsilon+P \right) \delta u_{i} \,, \\
\delta T^{ii} &= \left(\frac{\partial P}{\partial\mu}\right)_{T} \delta \mu + \left(\frac{\partial P}{\partial T}\right)_{\mu} \delta T - \eta (2\partial_{i} \delta u_{i}-\partial_{\gamma} \delta u_{\gamma}) \,, \,\,\,\,
\delta T^{ij} =-\eta (\partial_{j} \delta u_{i}+\partial_{i} \delta u_{j}) \,, \\
\delta J^{t} &= \left(\frac{\partial\rho}{\partial\mu}\right)_{T} \delta \mu + \left(\frac{\partial\rho}{\partial T}\right)_{\mu} \delta T \,,  \\ 
\delta J^{i} &= \rho \delta u_{i} + \sigma\left( -\partial_{i} \delta \mu  + \frac{\mu}{T} \partial_{i} \delta T  + H \varepsilon_{ij} \delta u_{j} \right) \,,
\end{split}
\end{align} 
where $\varepsilon_{ij}$ is the Levi-Civita symbol.
Plugging \eqref{FLUCFLUC} into the equations of motion \eqref{CONSERVATIONEQ} and also performing a Fourier transformation with the plane wave form $e^{-i \omega t + i k x}$, we obtain the four coupled equations:
\begin{align}\label{EOMSET1}
\begin{split}
 0 &= \omega \left[  \emu  \delta \mu +   \eT \delta T    \right] - k (\epsilon + P) \delta u_{x} \,, \\
 0 &= \omega (\epsilon + P) \delta u_{x} - k \left[  \Pmu  \delta \mu +   \PT \delta T    \right] + i k^2 \eta \, \delta u_{x} + i \sigma H^2 \delta u_{x} - i H \rho \delta u_{y} \,, \\
 0 &= \omega (\epsilon + P) \delta u_{y} + k H \sigma \left(\delta \mu - \frac{\mu}{T} \delta T\right) + i H \rho \delta u_{x} + i \sigma H^2 \delta u_{y} + i k^2 \eta \delta u_{y} \,, \\
 0 &= \omega \left[  \rhomu  \delta \mu +   \rhoT \delta T    \right] - k \rho \delta u_{x} - k \sigma H \delta u_{y} + i k^2 \sigma \left(\delta\mu - \frac{\mu}{T}\delta T \right) \,,
\end{split}
\end{align} 
which are consistent with equations in \cite{Hartnoll:2007ih} where the thermodynamic relation $\epsilon+P=s T +\mu \rho$ holds.
The equations of motion \eqref{EOMSET1} can also be expressed as the matrix form, $\mathcal{M}\cdot\mathcal{V}=0$, with
\begin{equation} \label{MATEOM}
\mathcal{M} :=
\begin{pmatrix}
            - k (\epsilon + P)   &  \omega \eT  &  0  & \omega  \emu \\
            \omega (\epsilon + P) + i k^2 \eta + i \sigma H^2  & -k \PT  & -i H \rho & -k \Pmu \\
            i H \rho & -\frac{k H \sigma \mu}{T} & \omega (\epsilon + P) + i k^2 \eta + i \sigma H^2 & k H \sigma \\
            -k \rho & \omega \rhoT -\frac{i k^2 \sigma \mu}{T}& -k \sigma H & \omega \rhomu + i k^2 \sigma
\end{pmatrix}
\end{equation}
and the vector $\mathcal{V}:=(\delta u_x, \, \delta T, \, \delta u_y, \, \delta \mu)$.
Then one can obtain the dispersion relations, $\omega=\omega(k)$, by the determinant of \eqref{MATEOM}:
\begin{align}\label{DETDISPER}
\begin{split}
0 \,=\, \text{det} \, \mathcal{M} \,:=\, \mathcal{M}_1 + \mathcal{M}_2 + \mathcal{M}_3 \,,
\end{split}
\end{align} 
where
\begin{align}\label{DETDISPER2}
\begin{split}
\mathcal{M}_1 &= \omega \rho H^2  \left[i k^2 \sigma (\epsilon+P) \left( \mu \rhomu + T \rhoT \right)  + \omega \rho T \left( \eT\rhomu -\emu\rhoT   \right)  \right]   \,, \\
\mathcal{M}_2 &= i k^2 T \sigma X (Z-i H^2 \sigma) + \omega T Z \left( X \rhomu - Y \rhoT \right) + i \frac{Z}{H^2 \sigma} \mathcal{M}_3 \,, \\
\mathcal{M}_3 &=  k^2 H^2 \sigma \left[ \mu \sigma Y - i \omega \rho T \left( \eT \Pmu - \emu \PT  \right) \right]  \,,
\end{split}
\end{align} 
with
\begin{align}\label{DETDISPER3}
\begin{split}
X &=  k^2 \left( \epsilon +P \right) \PT -\omega \,Z\, \eT  \,, \qquad 
Y =   k^2 \left( \epsilon +P \right) \Pmu -\omega \,Z\, \emu  \,, \\
Z &=   \omega \left( \epsilon +P \right) + i k^2 \eta + i \sigma H^2 \,.
\end{split}
\end{align} 
In the following sections, we study the dispersion relations by \eqref{DETDISPER}-\eqref{DETDISPER3} at zero density in Section \ref{ZDS} and at finite density in Section \ref{FDS}, respectively.
%In particular, we consider the following three cases in the main context
%1) (Zero density, Zero magnetic fields) 2) (Zero density, Finite magnetic fields) 3) (Finite density, Finite magnetic fields)

Note that the highest order of $\omega$ in \eqref{DETDISPER2} is $\mathcal{O}\left(\omega^4\right)$ from $\mathcal{M}_2$ so that one can solve \eqref{DETDISPER} to obtain $\omega(k)$ explicitly. However, the analytic expression of $\omega(k)$ is not so illuminating and complicated so we do not show it here. Instead, we will display its plots when we compare with quasi-normal modes from holography in the next section: see solid lines in Fig. \ref{QNMSFIG1} and Fig. \ref{QNMSFIG2}. Furthermore, for the analysis of hydrodynamic modes of $\omega(k)$, we will show the analytic expression of the dispersion relation at the small $k$ regime in the following subsections.

\subsection{Zero density}\label{ZDS}

Let us first consider the hydrodynamics with no density ($\rho=0$).\footnote{At zero density, the chemical potential is also vanishing, $\mu=0$.}
Moreover, motivated by $M2$-brane magneto-hydrodynamics~\cite{Buchbinder:2008dc}, we may set 
\begin{align}\label{ZDLIT}
\begin{split}
\rhoT= \emu = \Pmu = 0 \,,
\end{split}
\end{align} 
which will be verified by holography in the next section.

For the case of zero density with \eqref{ZDLIT}, one can check that $\mathcal{M}_1=\mathcal{M}_3=0$ in \eqref{DETDISPER2} so that \eqref{DETDISPER} becomes
\begin{align}\label{ZDD1}
\begin{split}
\quad 0 \,=\, \mathcal{M}_2  = X \left[ Z\left( i k^2 \sigma + \omega \rhomu \right) + k^2 H^2 \sigma^2 \right]\,,
\end{split}
\end{align} 
where $X$ and $Z$ are given in \eqref{DETDISPER3}. Note that $\mathcal{M}_2$ in \eqref{ZDD1} is decoupled into two parts: one from $X$ and its rest. This reflects the fact that the coupled equations in \eqref{EOMSET1} can be decoupled into two decoupled pairs at zero density~\cite{Buchbinder:2008dc,Buchbinder:2009aa,Buchbinder:2009mk}: i) ($\delta u_x, \, \delta T$) sector in \eqref{EOMSET1}, called the sound channel; ii) ($\delta u_y, \, \delta \mu$) sector in \eqref{EOMSET1}, called the shear channel.\footnote{Note that this decoupling can also be seen as a block-diagonalization in \eqref{MATEOM}.} In particular, the sound channel corresponds to $X$ in \eqref{ZDD1} and the shear channel comes in its rest.

\paragraph{Sound channel:}
In the sound channel, depending on $H$, one can have the following $\omega(k)$ in the small wave vector regime:
\begin{align}
%\begin{split}
(H=0)&: \quad \omega = \pm \sqrt{\frac{\partial P}{\partial \epsilon}} k \,-\, i  \frac{\eta}{2(\epsilon+P)} \,k^2  \,, \label{SCDSP1} \\
(H\neq0)&: \quad \omega = -i  \frac{\partial P}{\partial \epsilon} \frac{\epsilon + P}{\sigma H^2} \,k^2  \,, \qquad\quad \omega = -i \frac{\sigma H^2}{\epsilon+P} \,. \label{SCDSP2}
%\end{split}
\end{align} 
Thus, the sound mode \eqref{SCDSP1} at $H=0$ shows a drastic change into \eqref{SCDSP2} at finite $H$: the former is the energy diffusion mode and the later gapped mode is a damping frequency of the cyclotron mode~\cite{Hartnoll:2007ih,Hartnoll:2007ip}.\footnote{This change is due to the fact that the small $H$ limit does not commute with the hydrodynamic limit of small $\omega$ and $k$~\cite{Buchbinder:2008dc,Buchbinder:2009aa}.} Note that it was shown \cite{Jeong:2021zhz,Jeong:2021zsv} that dispersions \eqref{SCDSP1}-\eqref{SCDSP2} are matched with the quasi-normal modes in holography and  the lower/upper bound of the energy diffusion constant is investigated. See \cite{Jeong:2021zhz} to verify that the diffusion mode in \eqref{SCDSP2} corresponds to the energy diffusion.

\paragraph{Shear channel:}
Within the shear channel, similar to the sound channel, dispersions also depend on $H$ as follows:
\begin{align}
%\begin{split}
(H=0)&: \quad \omega = -i  \frac{\sigma}{\left(\frac{\partial\rho}{\partial\mu}\right)_{T}} \,k^2  \,, \qquad\quad\,\,\,
\omega = -i \frac{\eta}{\epsilon+P} k^2 \,, \label{MAGNETO1} \\
(H\neq0)&: \quad \omega = -i  \frac{\eta}{H^2 \left( \frac{\partial \rho}{\partial \mu} \right)_{T}} \,k^4  \,, \qquad \omega = -i \frac{\sigma H^2}{\epsilon+P} \,. \label{MAGNETO2}
%\end{split}
\end{align} 
At $H=0$, there are two gapless mode in \eqref{MAGNETO1}: the former is the charge diffusion mode and the other the shear diffusion mode. Furthermore, as in the sound channel, the shear channel has a gapless mode as well as  the cyclotron mode at $H\neq0$ in \eqref{MAGNETO2}: the gapless mode is called the sub-diffusion mode.\footnote{Considering the sub-leading order correction $\mathcal{O}(k^2)$, one can check that the gapped mode in \eqref{SCDSP2} is different from the one in \eqref{MAGNETO2}. See also \cite{Gromov:2020yoc} for the sub-diffusive modes within fracton hydrodynamics.} We will show dispersions \eqref{MAGNETO1}-\eqref{MAGNETO2} are consistent with quasi-normal modes in holography in the next section.

For a summary of the dispersion relations from hydrodynamics at zero density, \eqref{SCDSP1}-\eqref{MAGNETO2}, see Table. \ref{ST1}.
\begin{table}[]
\begin{tabular}{| C{2.7cm} | C{5.5cm} | C{5.5cm}  |}
\hline
             & $H=0$    & $H\neq0$  \\ 
 \hline
 \hline
             &  Sound mode \eqref{SCDSP1},             & Energy diffusion \eqref{SCDSP2}, \\
Gappless mode &  Charge diffusion mode \eqref{MAGNETO1}, &  Sub-diffusion mode \eqref{MAGNETO2}. \\ 
             &  Shear diffusion mode \eqref{MAGNETO1}. &    \\
 \hline
     Gapped mode    &   None    &        Cyclotron mode \eqref{SCDSP2}, \eqref{MAGNETO2}.          \\
 \hline
\end{tabular}
\caption{Summary of the dispersion relations from hydrodynamics at \textit{zero} density.}\label{ST1}
\end{table}

\subsection{Finite density}\label{FDS}

Next, let us study the dispersion relations at finite density ($\rho\neq0$) in which \eqref{ZDLIT} no longer holds. One can notice that $\mathcal{M}_i$ in \eqref{DETDISPER2} are all non-zero at finite density in general. In other words, the sound channel \eqref{SCDSP1}-\eqref{SCDSP2} are coupled with the shear channel \eqref{MAGNETO1}-\eqref{MAGNETO2} at finite density.

The aim of this subsection is to study how the dispersion, \eqref{SCDSP1}-\eqref{MAGNETO2}, are changed in the presence of a finite density. For this purpose, we analyze two cases, ($H=0$) and ($H\neq0$), separately at finite density, i.e., we may follow a parallel analysis as in the zero density case \eqref{SCDSP1}-\eqref{MAGNETO2}. Furthermore, for the case of $H=0$, one can find the simplified $\mathcal{M}_i$ even at finite density ($\mathcal{M}_1=\mathcal{M}_3=0$).

\paragraph{Zero magnetic field ($H=0$):}

For the case of $H=0$ at finite density, \eqref{DETDISPER} becomes
\begin{align}\label{ZMFD1}
\begin{split}
0 \,=\, \mathcal{M}_2  = Z \Bigg[  &X \left( i k^2 \sigma + \omega \rhomu  \right)  + \omega k^2 \rho \left( \Pmu \eT - \PT \emu \right) \\
&- \omega Y \rhoT -i \frac{\mu}{T} \left( \omega k^2 \sigma  \left(Z-i H^2 \sigma \right) \emu  - k^4 \sigma (\epsilon+P) \Pmu \right)   \Bigg]\,,
\end{split}
\end{align} 
where $X$, $Y$, and $Z$ are given in \eqref{DETDISPER3}. One can notice that \eqref{ZMFD1} becomes \eqref{ZDD1} together with \eqref{ZDLIT} at $\rho=H=0$. Similar to \eqref{ZDD1}, \eqref{ZMFD1} is also decoupled into two parts, one from $Z$ and its rest, which reflects that the coupled equations \eqref{EOMSET1} are decoupled into two sectors at $H=0$: i) ($\delta u_x, \, \delta T, \, \delta \mu$) sector; ii) ($\delta u_y$) sector.\footnote{At $(\rho=0, H=0)$, equations consist of three sectors: (i) $\delta \mu$ sector; (ii) ($\delta u_x, \, \delta T$) sector; (iii) $\delta u_y$ sector. For $(\rho=0, H\neq0)$, (i) is coupled to (iii) as in  \eqref{ZDD1}, while (i) is coupled to (ii) at $(\rho\neq0, H=0)$ \eqref{ZMFD1}. At ($\rho\neq0, H\neq0$), all sectors are coupled together.}

From \eqref{ZMFD1}, one can find the dispersions at leading order in small wave vector as
\begin{align}
\omega &= \pm \sqrt{\frac{-\PT \rhomu (\epsilon+P) + \rho \left( \PT\emu +\rhoT (\epsilon+P)  \right) - \rho^2 \eT }{\left( \emu \rhoT - \eT \rhomu \right)(\epsilon+P)}} \quad k   \,, \label{SCDSP1def} \\
%
%\begin{split}
\omega &=   \frac{-i\,(\epsilon+P)\left(T \PT+\mu \Pmu\right) \sigma}{ (\epsilon+P) \,T\,\PT \rhomu - T \Pmu \left( (\epsilon+P)\rhoT + \PT \emu - \Pmu \eT  \right) } \,k^2  \,, \label{MAGNETO1def} \\
\omega &= -i \frac{\eta}{\epsilon+P} k^2\,, \label{MAGNETO2def}
%\end{split}
%
\end{align} 
in which $Z$ in \eqref{ZMFD1} produces the shear diffusion mode \eqref{MAGNETO2def}. At zero density with \eqref{ZDLIT}, one can check that \eqref{SCDSP1def} reduces to \eqref{SCDSP1} and \eqref{MAGNETO1def}-\eqref{MAGNETO2def} become \eqref{MAGNETO1}.\footnote{One may also try to find the sub-leading correction $\mathcal{O}(k^2)$ in \eqref{SCDSP1def} in the presence of finite density, which becomes the attenuation constant in \eqref{SCDSP1} at vanishing density.}

\paragraph{Finite magnetic field ($H\neq0$):}

When the system has both a density and a magnetic field, we cannot find a simple equation for \eqref{DETDISPER}, such as \eqref{ZDD1} or \eqref{ZMFD1}, because all the equations are coupled, i.e., \eqref{DETDISPER} consists of all non-zero $\mathcal{M}_{i}$ given in \eqref{DETDISPER2}. For such a case, the corresponding dispersions at small wave vector are
\begin{align}
\omega &=    i\,\frac{ \, \rhomu (\epsilon+P)^2 \,\, \sigma}{T\left[ \emu\rhoT-\eT\rhomu \right]\left(\rho^2 + H^2 \sigma^2 \right)} \,\, k^2  \,,   \label{disfin1} \\
\omega &= -i  \frac{\eta}{H^2 \left( \frac{\partial \rho}{\partial \mu} \right)_{T}} \,k^4 \,,\qquad\qquad \omega = \pm \frac{H \rho}{\epsilon+P} -i \frac{\sigma H^2}{\epsilon+P}     \label{disfin2}  \,.
\end{align} 
Note that we find the prefactor $\rhomu$ of the diffusion mode \eqref{disfin1} in its numerator, which was not shown in \cite{Hartnoll:2007ih}.\footnote{We also correct the overall sign in all gapless mode in \eqref{disfin1}-\eqref{disfin2}.} We will show that this prefactor will be important to match with quasi-normal modes of dyonic black holes in the next section. Furthermore, note also that \eqref{disfin1} becomes the energy diffusion mode \eqref{SCDSP2} at zero density together with \eqref{ZDLIT} only when this prefactor is considered.\footnote{Also the thermodynamic relation $\epsilon+P=s T +\mu \rho$ is being used.}

For a summary of the dispersion relations from hydrodynamics at finite density, \eqref{SCDSP1def}-\eqref{disfin2}, see Table. \ref{ST2}.
\begin{table}[]
\begin{tabular}{| C{2.7cm} | C{5.5cm} | C{5.5cm}  |}
\hline
             & $H=0$    & $H\neq0$  \\ 
 \hline
 \hline
             &   Sound mode \eqref{SCDSP1def},             &   Diffusion mode \eqref{disfin1}, \\
Gappless mode &  Diffusion mode \eqref{MAGNETO1def}, &  Sub-diffusion mode \eqref{disfin2}. \\ 
             &  Shear diffusion mode \eqref{MAGNETO2def}. &    \\
 \hline
     Gapped mode    &   None    &        Cyclotron mode  \eqref{disfin2}.          \\
 \hline
\end{tabular}
\caption{Summary of the dispersion relations from hydrodynamics at \textit{finite} density.}\label{ST2}
\end{table}
Comparing Table. \ref{ST2} with Table. \ref{ST1}, one can notice three things about the finite density effect in dispersion relations.

First, the density does not generate new modes. In other words, the density only comes in the coefficients of dispersions such as the sound velocity of \eqref{SCDSP1def}, diffusion constants of \eqref{MAGNETO1def} and \eqref{disfin1}.
Second, the density does not change the functional form of the shear modes: i) shear diffusion \eqref{MAGNETO1}, \eqref{MAGNETO2def}; ii) sub-diffusion mode \eqref{MAGNETO2}, \eqref{disfin2}, i.e., the shear modes are intrinsic function for a density.
Third, the cyclotron mode \eqref{disfin2} gets its real part $\frac{H \rho}{\epsilon+P}$ due to the finite density, called the cyclotron frequency, which is consistent with the zero wave vector analysis~\cite{Hartnoll:2007ih,Hartnoll:2007ip}.

%%%%%%%%%%%%%%%%%%%%%%%%%%%%%%%%%%%%%%
%    Section: 
%%%%%%%%%%%%%%%%%%%%%%%%%%%%%%%%%%%%%%
%\section{Holography at finite magnetic fields}\label{}
\section{Quasi-normal modes in dyonic black holes}\label{SECMHDDIS333}

\subsection{Holographic setup}

We consider the dyonic black holes in (3+1) dimensions as
\begin{equation}\label{ACTIONH}
\begin{split}
S = \int \dd^4x \sqrt{-g} \,\left( R \,+\, 6 \,-\, \frac{1}{4} F^2 \right) \,,
\end{split}
\end{equation}
where $F=\dd A$ is the field strength of the gauge field $A$ and we set units such that the gravitational constant $16\pi G=1$ and the AdS radius $L=1$.

Within \eqref{ACTIONH}, we consider the following ansatz for the background
\begin{equation}\label{BGMET}
\begin{split}
\dd s^2 =  -f(r)\, \dd t^2 +  \frac{1}{f(r)} \, \dd r^2  + r^2 (\dd x^2 + \dd y^2) \,,\quad   A = A_t(r) \, \dd t -\frac{H}{2} y \,\dd x \,+\, \frac{H}{2} x \, \dd y \,,
\end{split}
\end{equation}
where $H$ is the magnetic field. The blackening factor $f(r)$ and the temporal component of the gauge field $A_t(r)$ are 
\begin{equation}\label{BCF}
\begin{split}
 f(r) &\,= r^2 - \frac{m_{0}}{r} \,+ \, \frac{\mu^2 r_{h}^2 + H^2}{4\,r^2} \,, \quad m_{0} = r_{h}^3\left( 1 +  \frac{\mu^2 r_{h}^2+H^2}{4\, r_{h}^4} \right) \,, \\
 A_{t}(r) &\,= \mu \left( 1- \frac{r_{h}}{r} \right)\,,
\end{split}
\end{equation}
where $\mu$ is the chemical potential, $r_{h}$ is the horizon radius. $m_{0}$ is determined by $f(r_{h})=0$.

Thermodynamic quantities~\cite{Hartnoll:2007ih,Hartnoll:2007ip,Kim:2015wba,Blake:2015hxa} including the temperature $T$ with the density $\rho$ read
\begin{align}\label{HAWKINGT}
\begin{split}
 T &\,=\, \frac{1}{4\pi} \left( 3\,r_{h} \,-\, \frac{\mu^2 r_{h}^2 + H^2}{4\,r_{h}^3}  \right) \,,  \quad \rho \,=\, \mu \, r_{h} \,, \quad s \,=\, 4\pi \, r_{h}^2  \,, \\
\epsilon &\,=\,  2r_{h}^3 + \frac{\mu^2 r_{h}}{2} + \frac{H^2}{2 r_{h}}  \,, \qquad\qquad P \,=\, r_{h}^3 + \frac{\mu^2 r_{h}}{4} - \frac{3 H^2}{4 r_{h}} \,,   
\end{split}
\end{align}
where $(s, \epsilon, P)$ are the entropy, energy and pressure density, respectively. Note that \eqref{HAWKINGT} satisfies the thermodynamic relation 
\begin{align}\label{SMARR}
 \epsilon + P \,=\, s \, T + \mu \, \rho \,,
\end{align}
Furthermore, using \eqref{HAWKINGT}, one can also find other thermodynamic quantities
\begin{align}\label{}
\begin{split}
 \rhomu  \,, \,\, \rhoT \,, \,\, \emu \,, \,\, \eT \,, \,\,  \Pmu \,, \,\,  \PT \,, 
\end{split}
\end{align}
which are non-vanishing functions in terms of ($T, \, \mu, \, H$) in general. However, one can easily check that some of them, \eqref{ZDLIT}, could be zero at $\rho=\mu=0$.

\subsection{Fluctuations and the determinant method}\label{DTMF}

In order to study quasi-normal modes of dyonic black holes \eqref{ACTIONH}, we consider the fluctuations $\delta g_{\mu\nu}$ and $\delta A_{\mu}$ 
\begin{align}\label{}
\begin{split}
g_{\mu\nu} \,\rightarrow\, g_{\mu\nu} + \delta g_{\mu\nu} \,, \quad A_{\mu} \,\rightarrow\, A_{\mu} + \delta A_{\mu} \,,
\end{split}
\end{align} 
where $g_{\mu\nu}$ and $A_{\mu}$ are the background fields \eqref{BGMET}. 
To proceed, it is convenient to consider the radial gauge
\begin{align}\label{}
\begin{split}
\delta g_{tr} = \delta g_{rr} =\delta g_{xr}=\delta g_{yr} =0  \,, \quad \delta A_{r} = 0 \,.
\end{split}
\end{align} 
In order to be consistent with the hydrodynamics given in previous section, we also consider all fluctuations to be functions of $(t, r, x)$, i.e., 
\begin{align}\label{FLUCOURSETUP}
\begin{split}
\delta g_{\mu\nu} &= h_{\mu\nu}(r) \,e^{-i \, \omega \, t + i \, k \, x} \,,\, \quad  \delta A_{\mu} = a_{\mu}(r) \,e^{-i \, \omega \, t + i \, k \, x} \,.
\end{split}
\end{align} 

\paragraph{Equations of motion for quasi-normal modes:}
Using \eqref{FLUCOURSETUP}, at the linearized fluctuation level of the Einstein equations and Maxwell equations, one can find \textit{nine} second-order equations and \textit{five} first-order constraints.
This implies that there are \textit{four} independent fluctuations associated with the diffeomorphism invariance together with the gauge invariance~\cite{Kovtun:2005ev}.
We find them to be
\begin{align}\label{GIVOUR}
\begin{split}
Z_{H_1} &\,:=\, k \,h_{t}^{y} \,+\, \omega \, h_{x}^{y}  \,, \\
Z_{A_1} &\,:=\, k \,a_{t} \,+\, \omega \, a_{x} \,-\, \frac{i H\,\omega}{k} h_{x}^{y} \,-\,\frac{k\,r}{2} \,A_{t}' \, h_{y}^{y} \,, \\
Z_{H_2} &\,:=\, \frac{4 k}{\omega} \, h_{t}^{x} \,+\,  2 h_{x}^{x} - \left( 2 - \frac{k^2}{\omega^2}\frac{f'(r)}{r} \right) h_{y}^{y}  + \frac{2k^2}{\omega^2}\frac{f(r)}{r^2} h_{t}^{t}  \,, \\
Z_{A_2} &\,:=\, a_{y}  \,+\, \frac{i H}{2k} \left(h_{x}^{x}-h_{y}^{y}\right) \,,
\end{split}
\end{align}
in which the index of the metric fluctuation is raised with \eqref{BGMET}.
Note that \eqref{GIVOUR} is consistent with \cite{Herzog:2002fn,Edalati:2010pn,Edalati:2010hk,Davison:2011uk,Davison:2013bxa,Davison:2013jba} at $(H=0, \, A_t\neq0)$, and \cite{Buchbinder:2008dc,Buchbinder:2009aa,Buchbinder:2009mk,Jeong:2021zhz,Jeong:2021zsv} at $(H\neq0, \, A_t=0)$.
To our knowledge, the independent fluctuation variables \eqref{GIVOUR} in the presence of both $H$ and $A_t$ (or $\mu$) was not shown in previous literature. 
Also note that, at zero density ($A_t=0$), the fluctuation variables \eqref{GIVOUR} can be decomposed into two sectors: i) $(Z_{H_1}, \, Z_{A_1})$; \, ii) $(Z_{H_2}, \, Z_{A_2})$. In the field theory language given in section \ref{ZDS}, the former one corresponds the shear channel and the other is the sound channel.

Then, one can obtain four second-order equations for \eqref{GIVOUR} in the following form:
\begin{align}\label{ZAZHEOM}
\begin{split}
&0 \,=\, Z_{H_1}''  \,+\, \sum a_{i}^{A} \,Z_{A_i}'    \,+\, \sum a_{i}^{H} \,Z_{H_i}'   \,+\, \sum \tilde{a}_{i}^{A}  \,Z_{A_i}  \,+\,  \sum \tilde{a}_{i}^{H} \,Z_{H_i}      \,, \\
&0 \,=\, Z_{A_1}''  \,\,+\, \sum b_{i}^{A} \,Z_{A_i}'    \,+\, \sum b_{i}^{H} \,Z_{H_i}'   \,\,+\, \sum \tilde{b}_{i}^{A}  \,Z_{A_i}  \,+\,  \sum \tilde{b}_{i}^{H} \,Z_{H_i}      \,, \\
&0 \,=\, Z_{H_2}''  \,\,+\, \sum c_{i}^{A} \,Z_{A_i}'    \,+\, \sum c_{i}^{H} \,Z_{H_i}'   \,\,+\, \sum \tilde{c}_{i}^{A}  \,Z_{A_i}  \,+\,  \sum \tilde{c}_{i}^{H} \,Z_{H_i}     \,, \\
&0 \,=\, Z_{A_2}''  \,\,+\, \sum d_{i}^{A} \,Z_{A_i}'    \,+\, \sum d_{i}^{H} \,Z_{H_i}'   \,+\, \sum \tilde{d}_{i}^{A}  \,Z_{A_i}  \,+\,  \sum \tilde{d}_{i}^{H} \,Z_{H_i}       \,.
\end{split}
\end{align}
Since the coefficients of \eqref{ZAZHEOM}, $\left(a_{i}^{A, H}, \, \tilde{a}_{i}^{A, H}, \, b_{i}^{A, H}, \,  \tilde{b}_{i}^{A, H}, \, c_{i}^{A, H}, \,  \tilde{c}_{i}^{A, H}, \,  d_{i}^{A, H}, \,  \tilde{d}_{i}^{A, H} \right)$, are lengthy and not illuminating we will not write them in the paper.

\paragraph{Determinant method:}

Next, we solve the equations \eqref{ZAZHEOM} with the boundary conditions: one from the horizon and the other at the AdS boundary.
Near the horizon ($r\rightarrow r_{h}$), the variables \eqref{GIVOUR} behave as
\begin{align}\label{APPENHORIZON}
\begin{split}
Z_{H_{i}} = (r-r_{h})^{\nu_{\pm}} \left( Z_{H_{i}}^{(0)} \,+\, Z_{H_{i}}^{(1)} (r-r_{h}) \,+\, \dots   \right ) \,, \\
Z_{A_{i}} = (r-r_{h})^{\nu_{\pm}} \left( Z_{A_{i}}^{(0)} \,+\, Z_{A_{i}}^{(1)} (r-r_{h}) \,+\, \dots   \right ) \,,
\end{split}
\end{align}
where $\nu_{\pm}:= \pm i\omega/4 \pi T$ and we choose $\nu_{-}$ which satisfies the incoming boundary condition at the horizon. Plugging \eqref{APPENHORIZON} into equations \eqref{ZAZHEOM}, one can check that higher-order horizon coefficients are determined by four independent horizon variables: $\left(Z_{H_{i}}^{(0)}, \,Z_{A_{i}}^{(0)}\right)$.

Near the AdS boundary ($r\rightarrow \infty$), the variables \eqref{GIVOUR} are expanded as
\begin{align}\label{BD2}
\begin{split}
&Z_{H_{i}} = Z_{H_{i}}^{(S)} \, r^{0} \,(1 \,+\, \dots) \,+\, Z_{H_{i}}^{(R)} \, r^{-3} \,(1 \,+\, \dots) \,, \\
&Z_{A_{i}} = Z_{A_{i}}^{(S)} \, r^{0} \,(1 \,+\, \dots) \,+\, Z_{A_{i}}^{(R)}\, r^{-1} \,(1 \,+\, \dots) \,,
\end{split}
\end{align}
where the superscripts denote that $(S)$ is the source and $(R)$ is the response term according to the holographic dictionary. 

Then, employing the determinant method~\cite{Kaminski:2009dh}, we can compute the quasi-normal modes. 
In particular, solving equations \eqref{ZAZHEOM} together with boundary conditions \eqref{APPENHORIZON}-\eqref{BD2}, one can construct the matrix of the sources, $S$-matrix, as follows:
\begin{align}\label{APPENSMATA}
\begin{split}
S = \left( 
\begin{array}{cccc} 
Z_{H_1}^{(S)(I)} & Z_{H_1}^{(S)(II)} & Z_{H_1}^{(S)(III)} & Z_{H_1}^{(S)(IV)} \\ [6pt]
Z_{A_1}^{(S)(I)} & Z_{A_1}^{(S)(II)} & Z_{A_1}^{(S)(III)} & Z_{A_1}^{(S)(IV)} \\ [6pt]
Z_{H_2}^{(S)(I)} & Z_{H_2}^{(S)(II)} & Z_{H_2}^{(S)(III)} & Z_{H_2}^{(S)(IV)} \\ [6pt]
Z_{A_2}^{(S)(I)} & Z_{A_2}^{(S)(II)} & Z_{A_2}^{(S)(III)} & Z_{A_2}^{(S)(IV)}
\end{array}  \right) \,.
\end{split}
\end{align}
Note that the $S$-matrix is a $4\times4$ matrix composed of four independent shooting variables at the horizon \eqref{APPENHORIZON}. Note also that $I (II,\, III,\, IV)$ in \eqref{APPENSMATA} means that the source terms are evaluated by the $I (II,\, III,\, IV)$-th shooting.
Finally, the dispersion relations, $\omega(k)$, of the dyonic black holes \eqref{ACTIONH} can be obtained by the value of ($\omega, k$) at which the determinant of the $S$-matrix \eqref{APPENSMATA} vanishes~\cite{Kaminski:2009dh}.

\subsection{Quasi-normal modes and hydrodynamics}

\paragraph{Transport coefficients in holography:}
In order to compare quasi-normal modes with the dispersion relations from hydrodynamics in the previous section, we need to identity the transport coefficients ($\sigma\,,\eta$) in addition to thermodynamic quantities \eqref{HAWKINGT}, which read 
\begin{align}\label{etasigma}
\begin{split}
\sigma = \left(\frac{s T}{\epsilon+P}\right)^2  \,, \quad \eta = \frac{s}{4\pi} \,,
\end{split}
\end{align}
where the conductivity $\sigma$ is given in \cite{Hartnoll:2007ih,Hartnoll:2007ip,Kim:2015wba,Blake:2015hxa}.\footnote{\label{ft14}See also \cite{Amoretti:2021fch,Amoretti:2020mkp,Amoretti:2019buu} for the recent development of magneto-transport properties in which the magnetic field is no longer taken to be of order one in derivatives.}
The shear viscosity $\eta$ in \eqref{etasigma} implies that the KSS bound~\cite{Kovtun:2004de,Kovtun:2003wp} is not violated in the presence of both a density and a magnetic field.

The shear viscosity can be evaluated holographically from the low frequency behavior of the shear correlator in the standard way~\cite{Davison:2015taa,Lucas:2015vna,Hartnoll:2016tri} \footnote{See also \cite{Alberte:2017oqx,Alberte:2017cch,Baggioli:2018bfa,Amoretti:2019cef} for the case with spontaneous symmetry breaking and \cite{Alberte:2015isw,Alberte:2016xja,Burikham:2016roo,Ciobanu:2017fef} for the explicit breaking case.} where the shear correlator can be computed from the shear equation at zero wave vector. One can easily check that the shear equation of the dyonic black hole \eqref{ACTIONH} with the background \eqref{BGMET} is 
\begin{align}\label{vm2ds}
\begin{split}
\left[ r^2 f(r) h_{x}^{y}{'}(r)  \right]^{'} + \omega^2 \frac{r^2}{f(r)} h_{x}^{y}(r) -M^2(r) h_{x}^{y}=0\,, \quad M^2(r)=0 \,,
\end{split}
\end{align}
where $M^2$ is the effective graviton mass. The vanishing graviton mass in \eqref{vm2ds} implies that the KSS bound is not violated~\cite{Kovtun:2003wp,Iqbal:2008by,Hartnoll:2016tri} so that $\eta$ is \eqref{etasigma}.\footnote{For the higher dimensional case~\cite{Jain:2015txa,Finazzo:2016mhm,Rebhan:2011vd,Giataganas:2013lga}, the KSS bound can be violated at finite magnetic fields.}

\paragraph{Quasi-normal modes at zero density:}
Then, using the determinant method, the thermodynamic quantities \eqref{HAWKINGT}, and the transport coefficients \eqref{etasigma}, one can compute the quasi-normal modes of the dyonic black holes and compare them with dispersion relations from hydrodynamics. 

\begin{figure}[]
\centering
     \subfigure[Re($\omega$) at $H/T^2=0$]
     {\includegraphics[width=4.83cm]{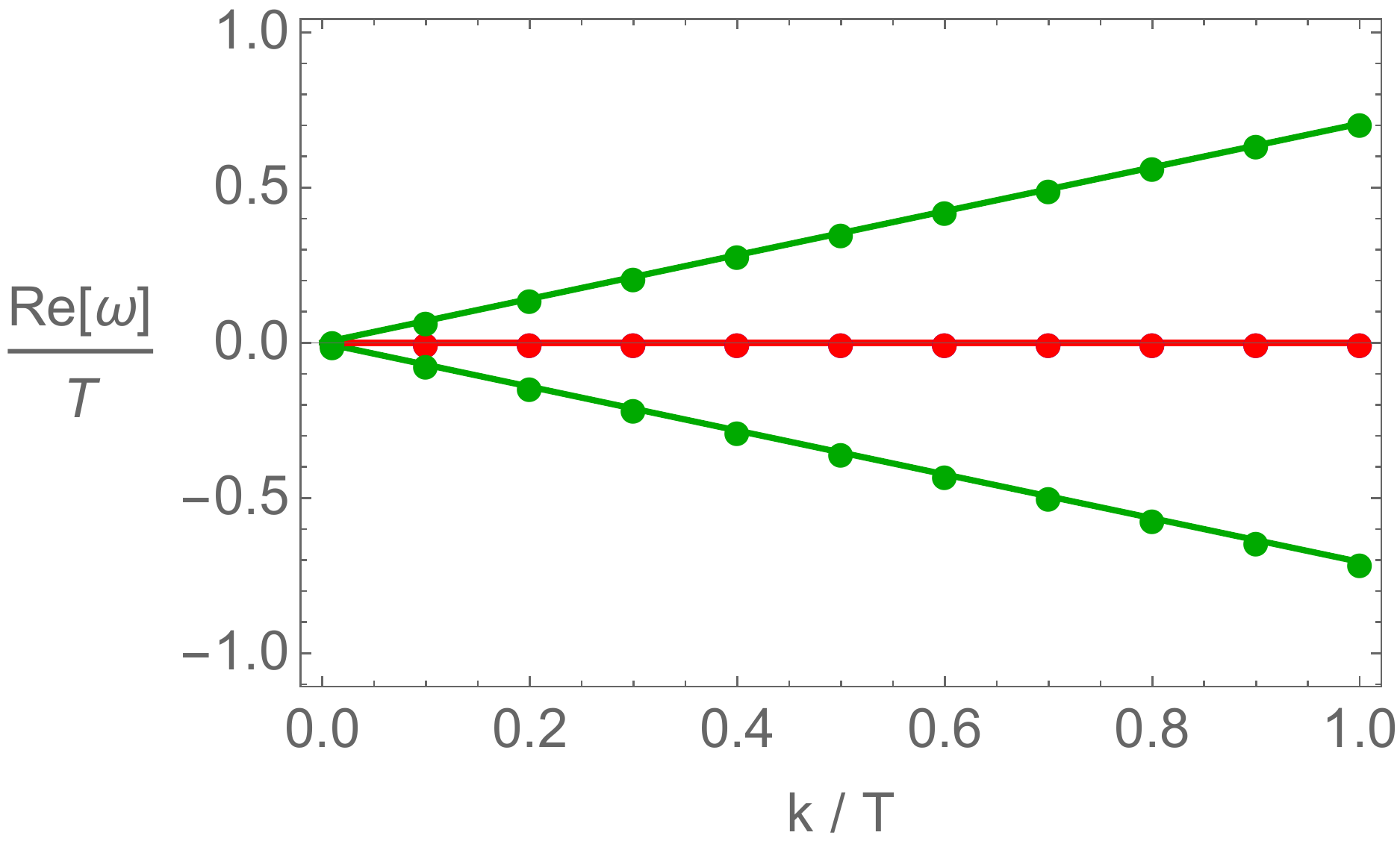} \label{}}
     \subfigure[Re($\omega$) at $H/T^2=5$]
     {\includegraphics[width=4.83cm]{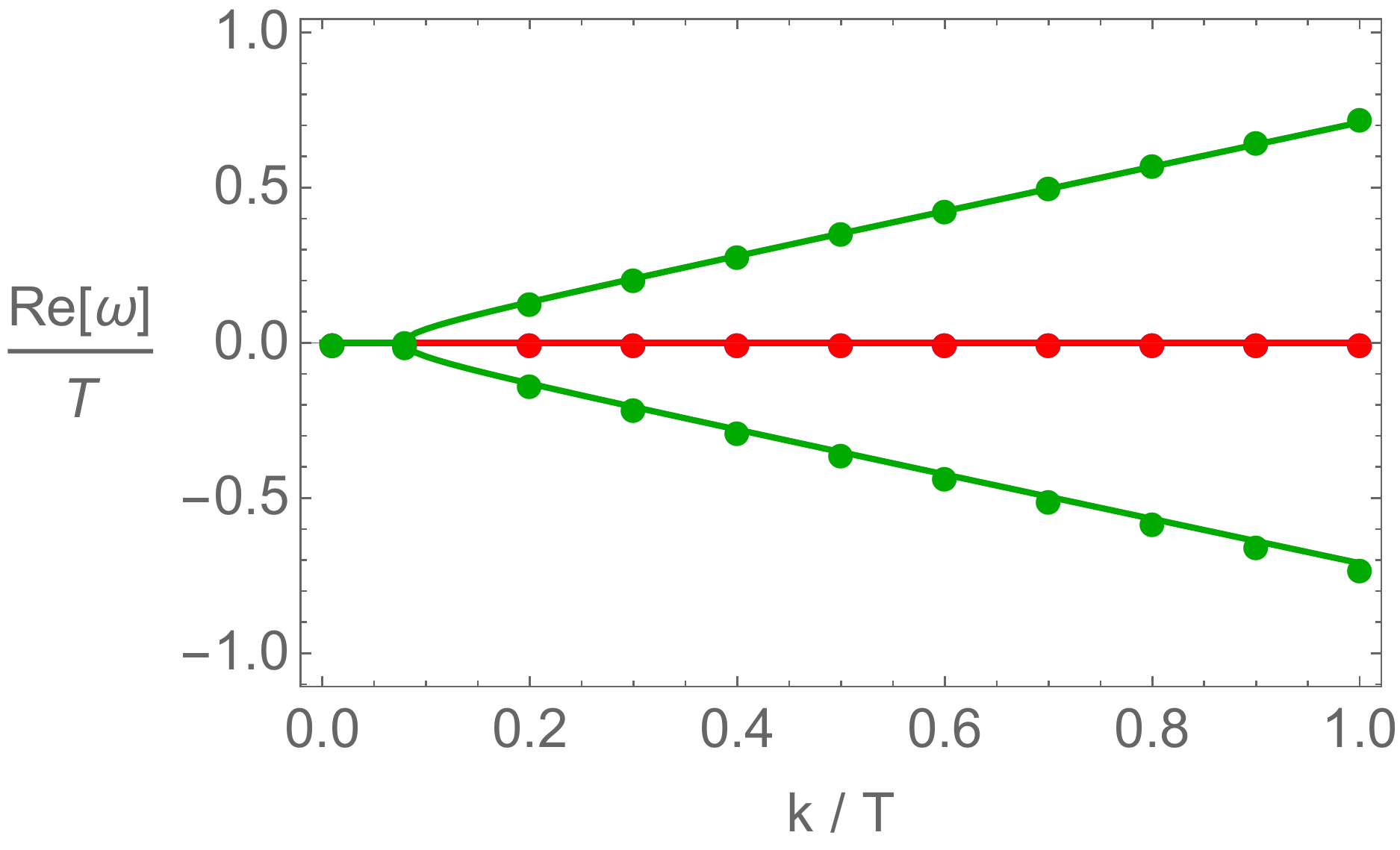} \label{}}
     \subfigure[Re($\omega$) at $H/T^2=10$]
     {\includegraphics[width=4.83cm]{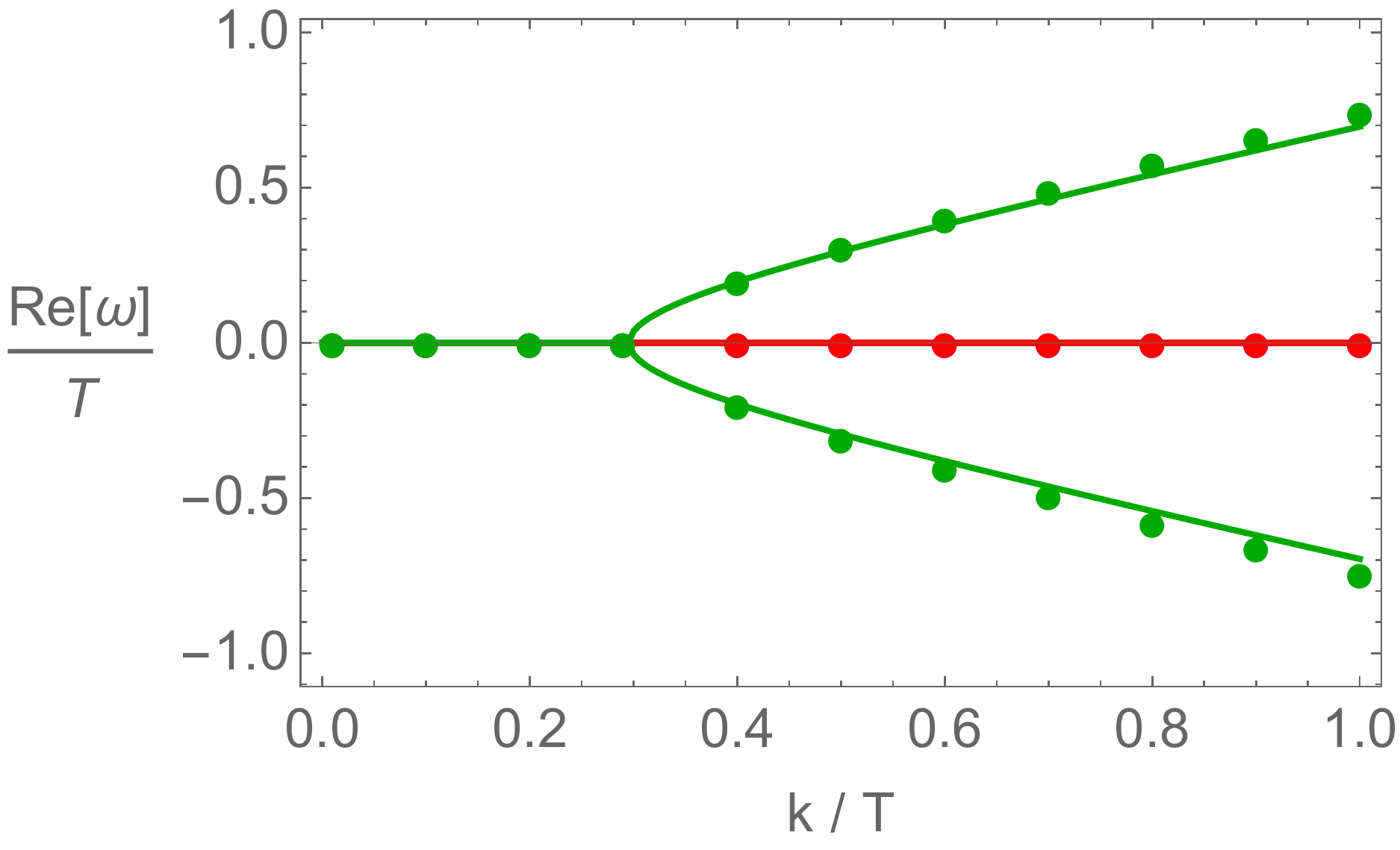} \label{}}
     
     \subfigure[Im($\omega$) at $H/T^2=0$]
     {\includegraphics[width=4.83cm]{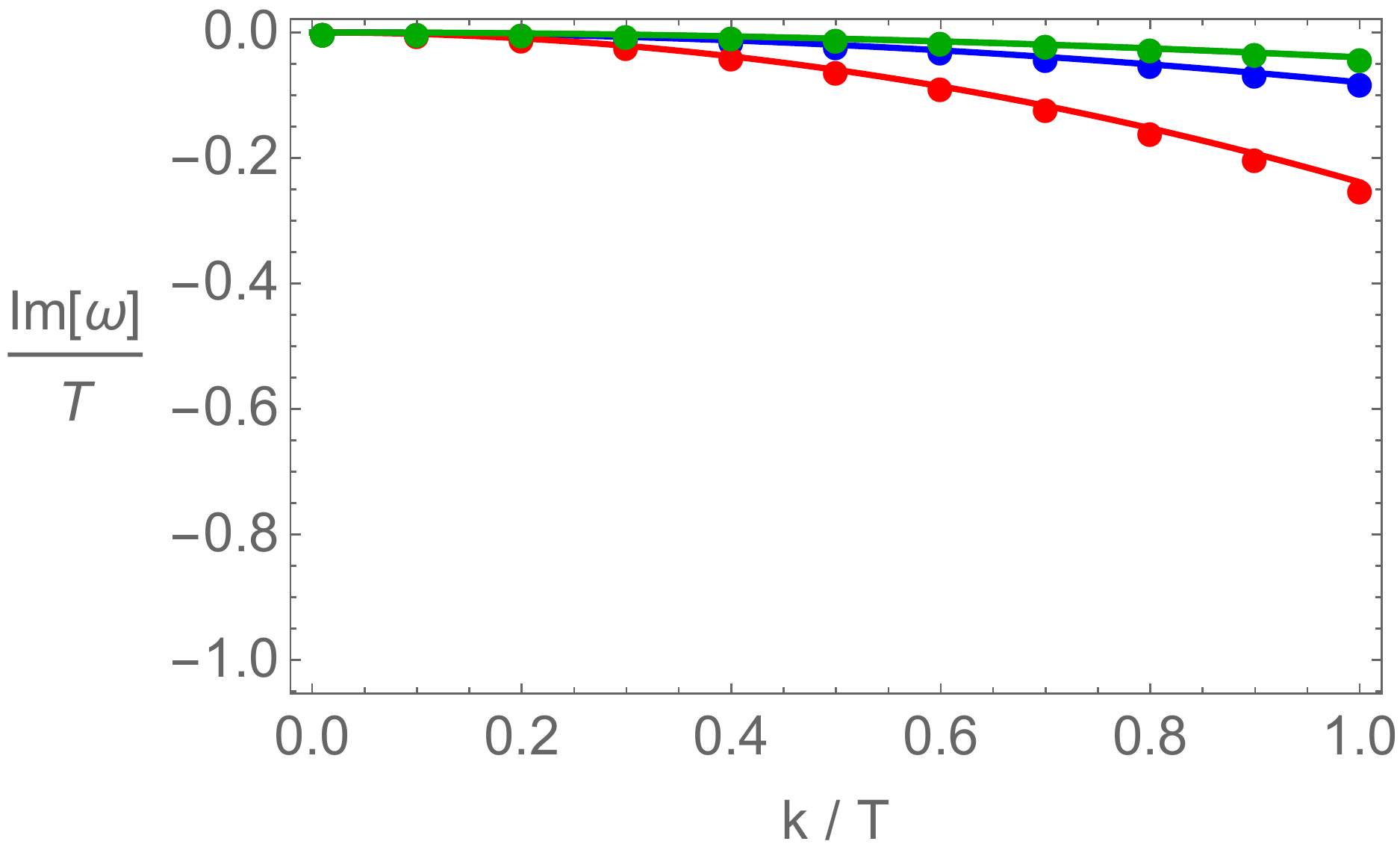} \label{}}
     \subfigure[Im($\omega$) at $H/T^2=5$]
     {\includegraphics[width=4.83cm]{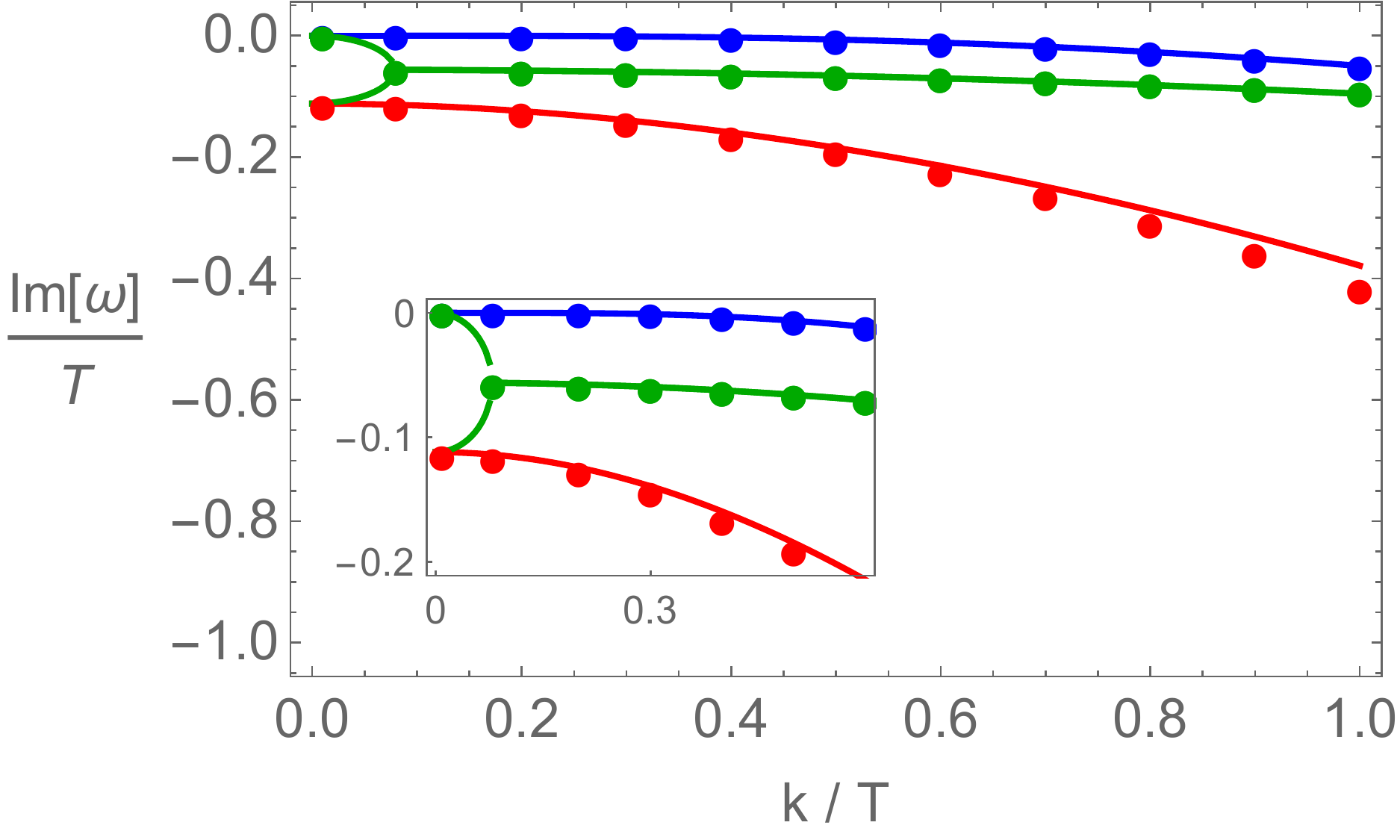} \label{}}
     \subfigure[Im($\omega$) at $H/T^2=10$]
     {\includegraphics[width=4.83cm]{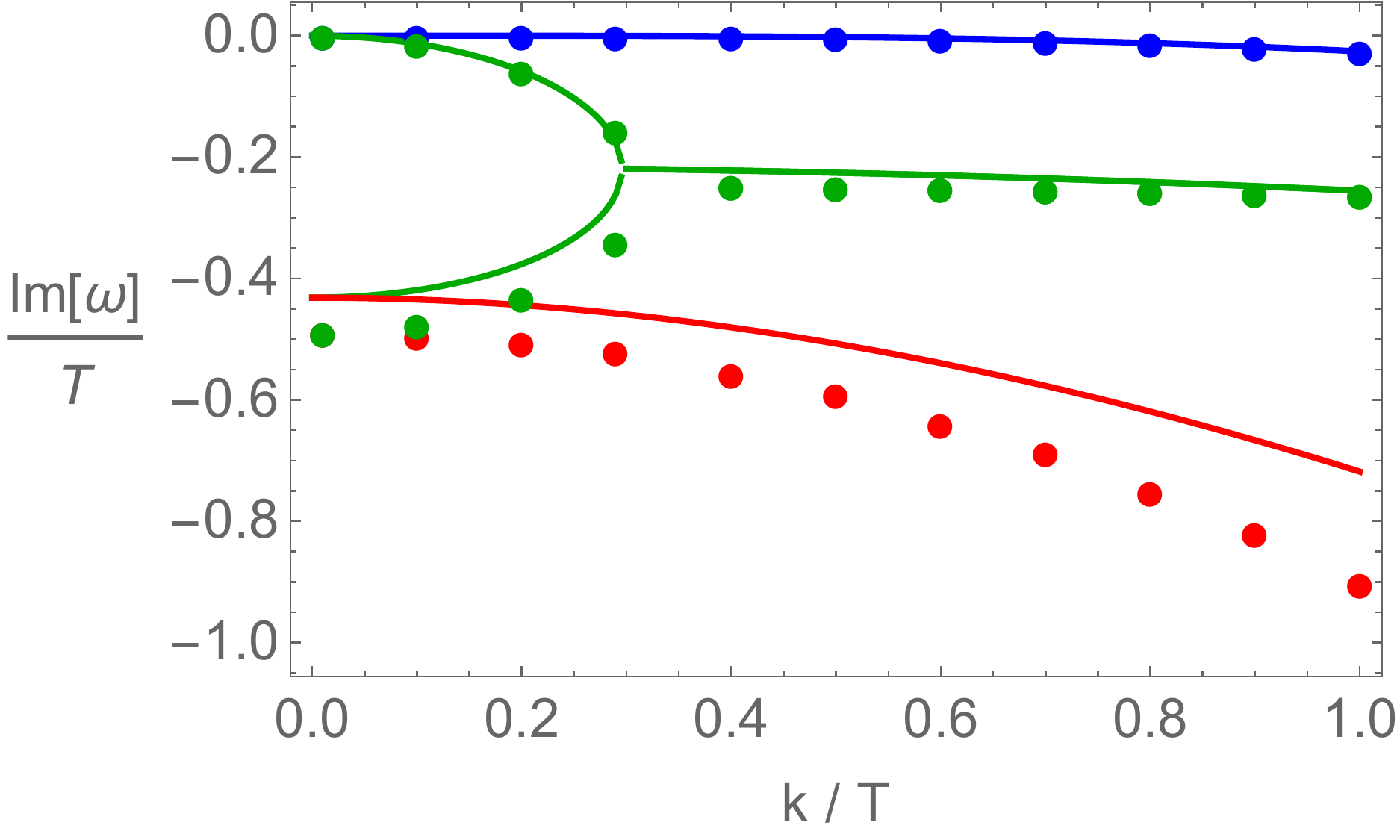} \label{}}
 \caption{Quasi-normal modes vs dispersion relations from hydrodynamics at \textit{zero} density ($\mu/T=0$). \textbf{Left:} (a) and (d) at $H/T^2=0$. \textbf{Center:} (b) and (e) at $H/T^2=5$. \textbf{Right:} (c) and (f) at $H/T^2=10$. \textbf{All figures:} colored dots are numerically computed quasi-normal modes and solid lines are dispersion relations from hydrodynamics by \eqref{DETDISPER}.}\label{QNMSFIG1}
\end{figure}

In Fig. \ref{QNMSFIG1}, we first display the quasi-normal modes at zero density ($\mu/T=0$) together with the dispersion relations from hydrodynamics: see also Table. \ref{ST1}.
For the $H=0$ case, (a) and (d), the green data corresponds to the sound mode \eqref{SCDSP1}, the red data is the charge diffusion mode \eqref{MAGNETO1}, and the blue data is the shear diffusion mode \eqref{MAGNETO1}.

For the finite $H$ case, (b) and (e) (or (c) and (f)), the green data consists two dispersions \eqref{SCDSP2}: the energy diffusion mode (gapless mode), the cyclotron mode (gapped mode). The red data is another cyclotron mode \eqref{MAGNETO2} and the blue data is the sub-diffusion mode \eqref{MAGNETO2}.

Note that quasi-normal modes have the deviation from dispersion relations of hydrodynamics as the magnetic field increases, e.g., see the cyclotron mode (green or red) in (f). This implies that dispersion relations of hydrodynamics is supposed to be valid in the coherent regime in which the momentum dissipation rate $\Gamma$ (the damping frequency in cyclotron mode \eqref{MAGNETO2}) is small as $\Gamma/T\ll1$ (or $H/T\ll1$)~\cite{Hartnoll:2007ih,Hartnoll:2007ip,Kim:2015wba,Blake:2015hxa}\footnote{Thus, we consider all the hydrodynamic dispersion relations in section \ref{SECMHDDIS} to be only valid at small magnetic fields. This may also imply that we assume the corrections in the thermodynamics due to the magnetic field is ignored in the HKMS magneto-hydrodynamics given in this paper. See also footnote \eqref{ft14}.}: the same argument also applies to the case where the energy diffusion mode appears due to the scalar (axion) field~\cite{Davison:2014lua}, i.e., $m/T\ll1$, $m$ is the coefficient from the scalar field.
Note also that the red and green data in Fig. \ref{QNMSFIG1} are the reproduction of \cite{Jeong:2021zhz}.

\begin{figure}[]
\centering
     \subfigure[Re($\omega$) at $H/T^2=0$]
     {\includegraphics[width=4.83cm]{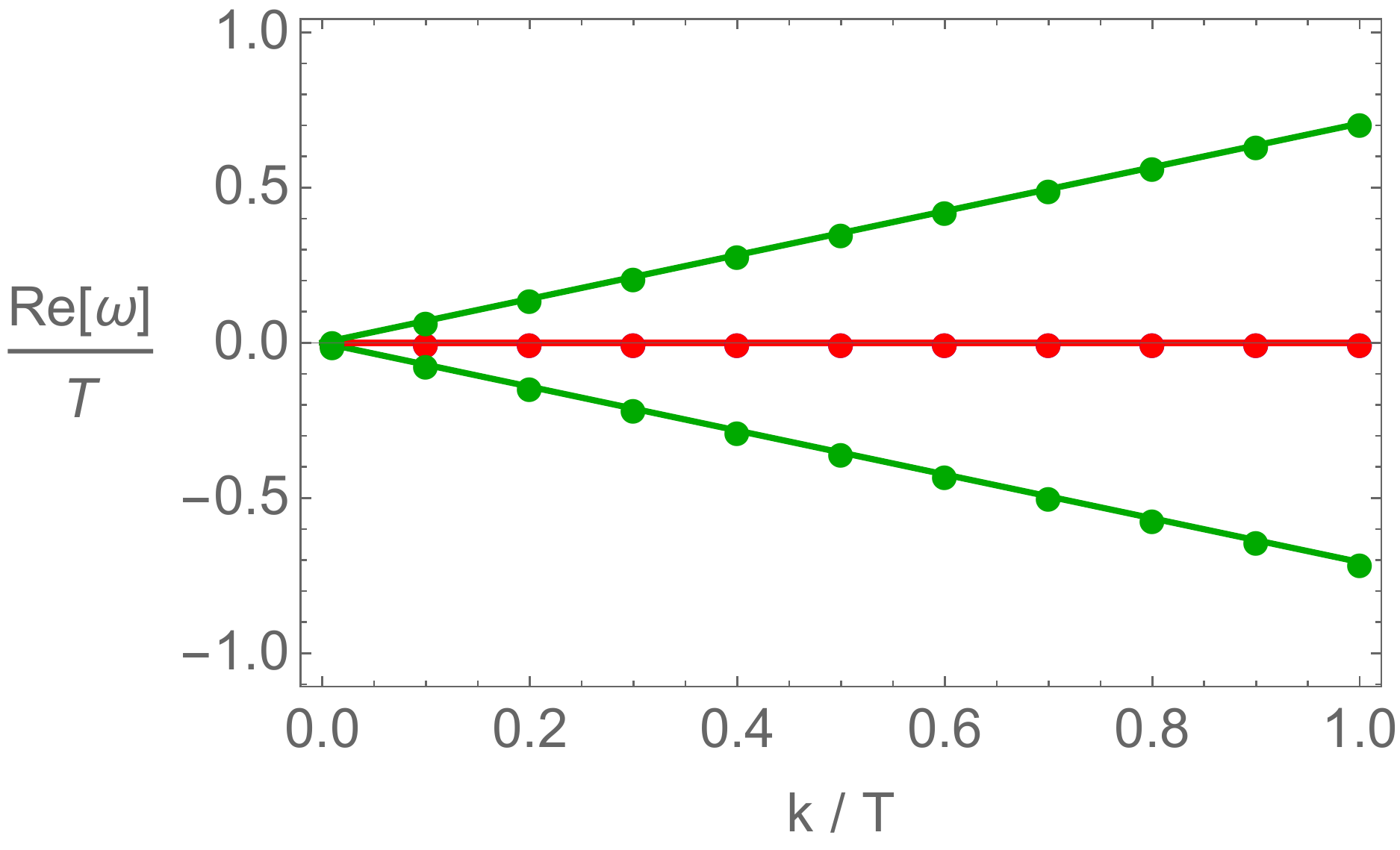} \label{}}
     \subfigure[Re($\omega$) at $H/T^2=5$]
     {\includegraphics[width=4.83cm]{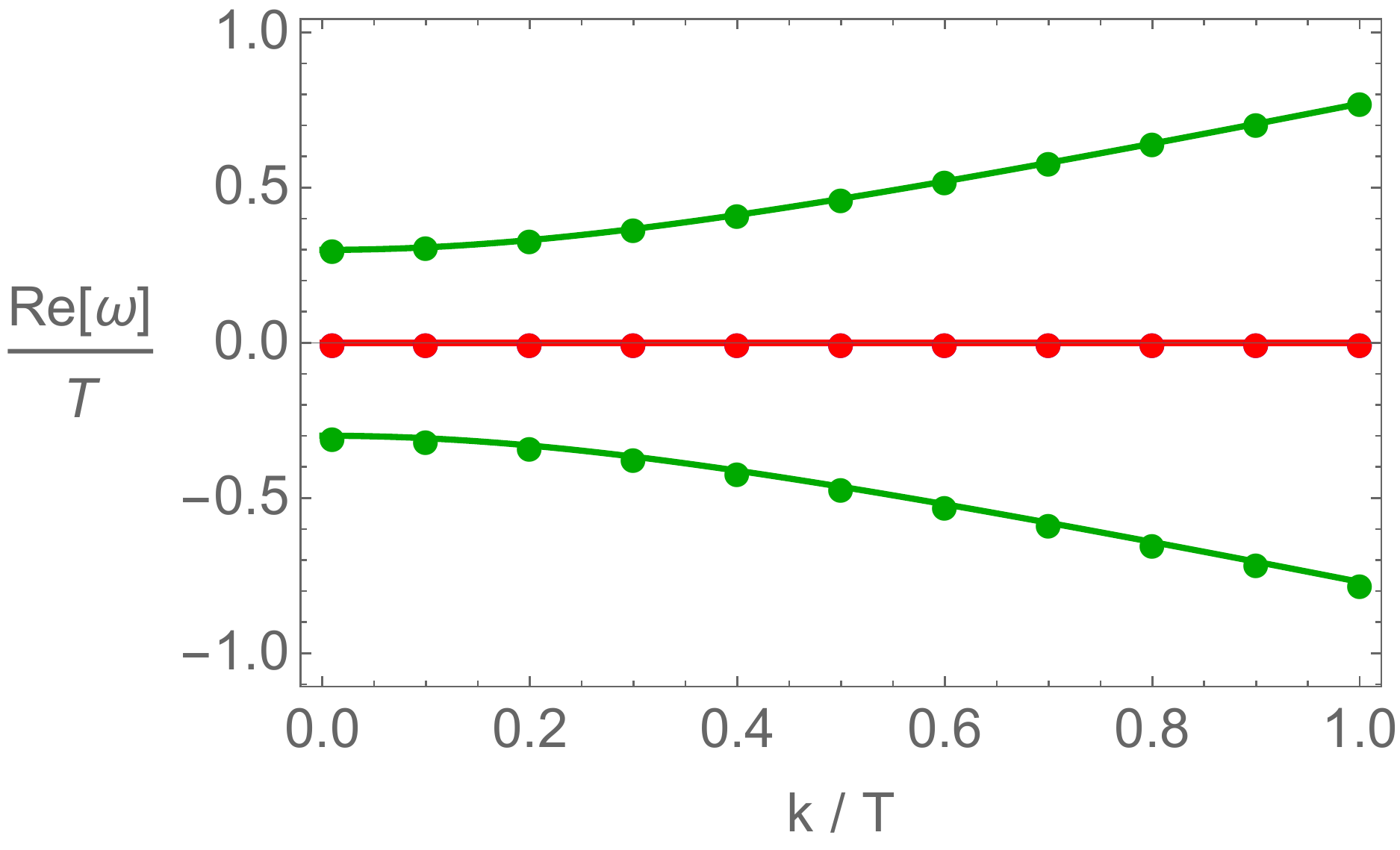} \label{}}
     \subfigure[Re($\omega$) at $H/T^2=10$]
     {\includegraphics[width=4.83cm]{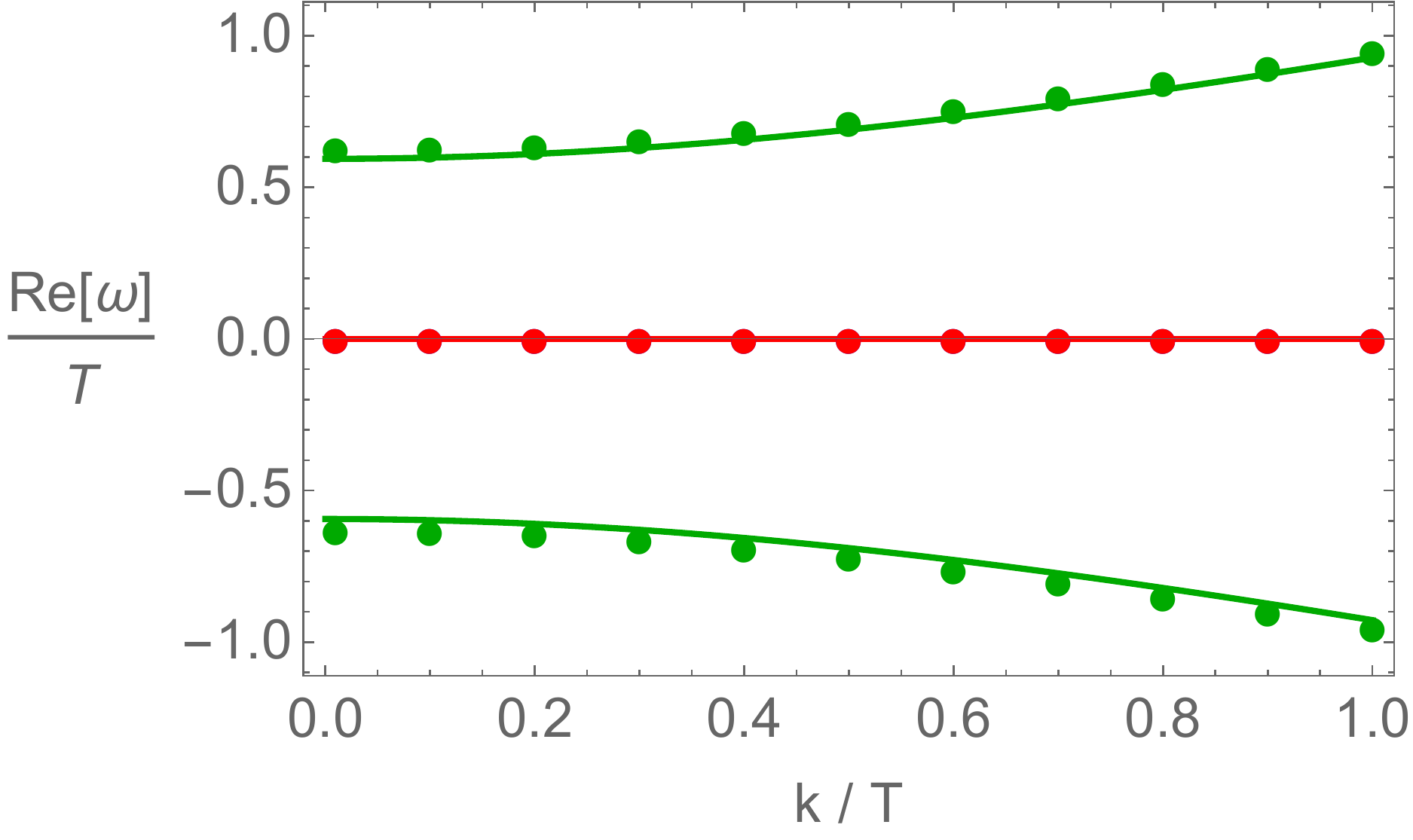} \label{}}
     
     \subfigure[Im($\omega$) at $H/T^2=0$]
     {\includegraphics[width=4.83cm]{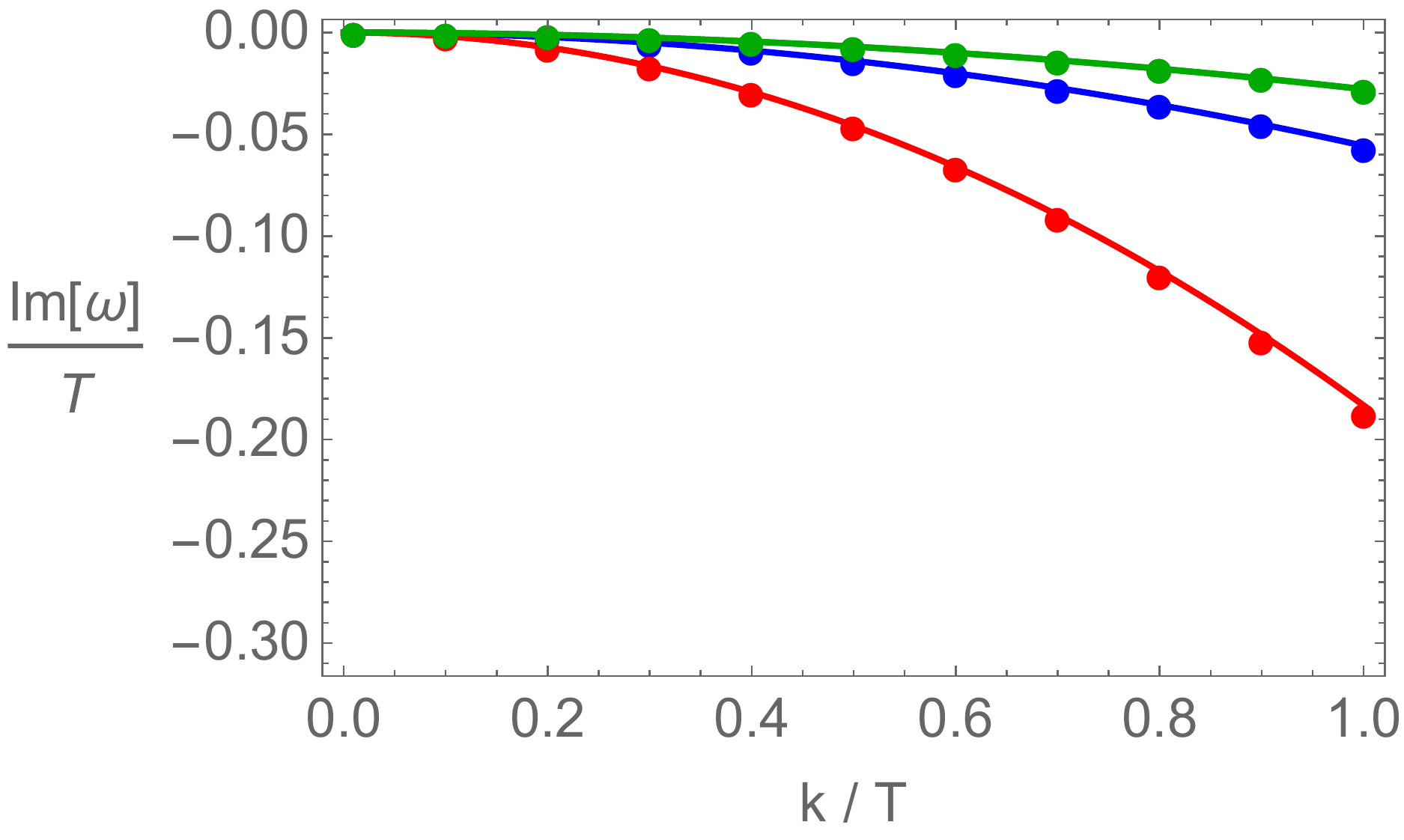} \label{}}
     \subfigure[Im($\omega$) at $H/T^2=5$]
     {\includegraphics[width=4.83cm]{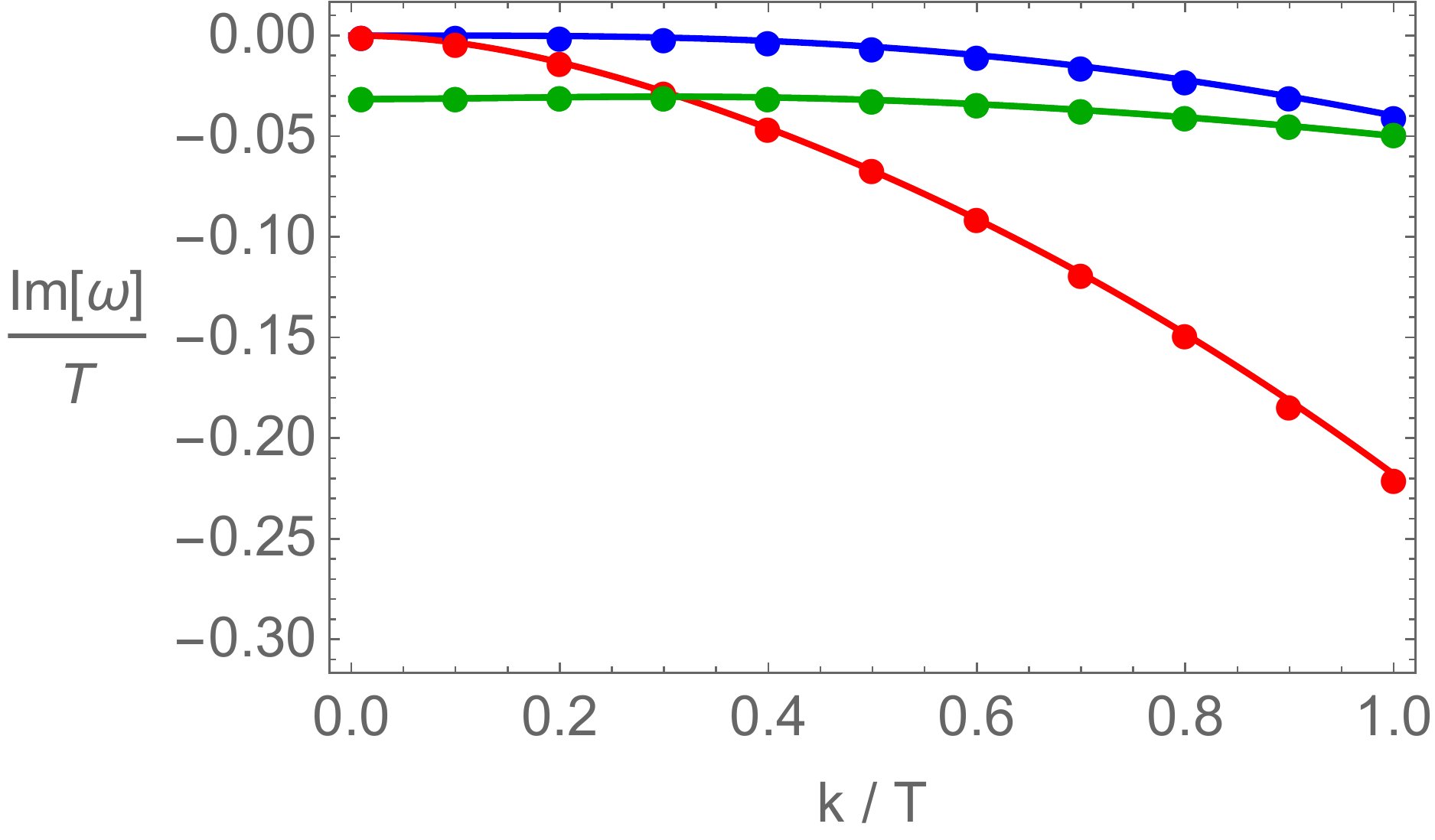} \label{}}
     \subfigure[Im($\omega$) at $H/T^2=10$]
     {\includegraphics[width=4.83cm]{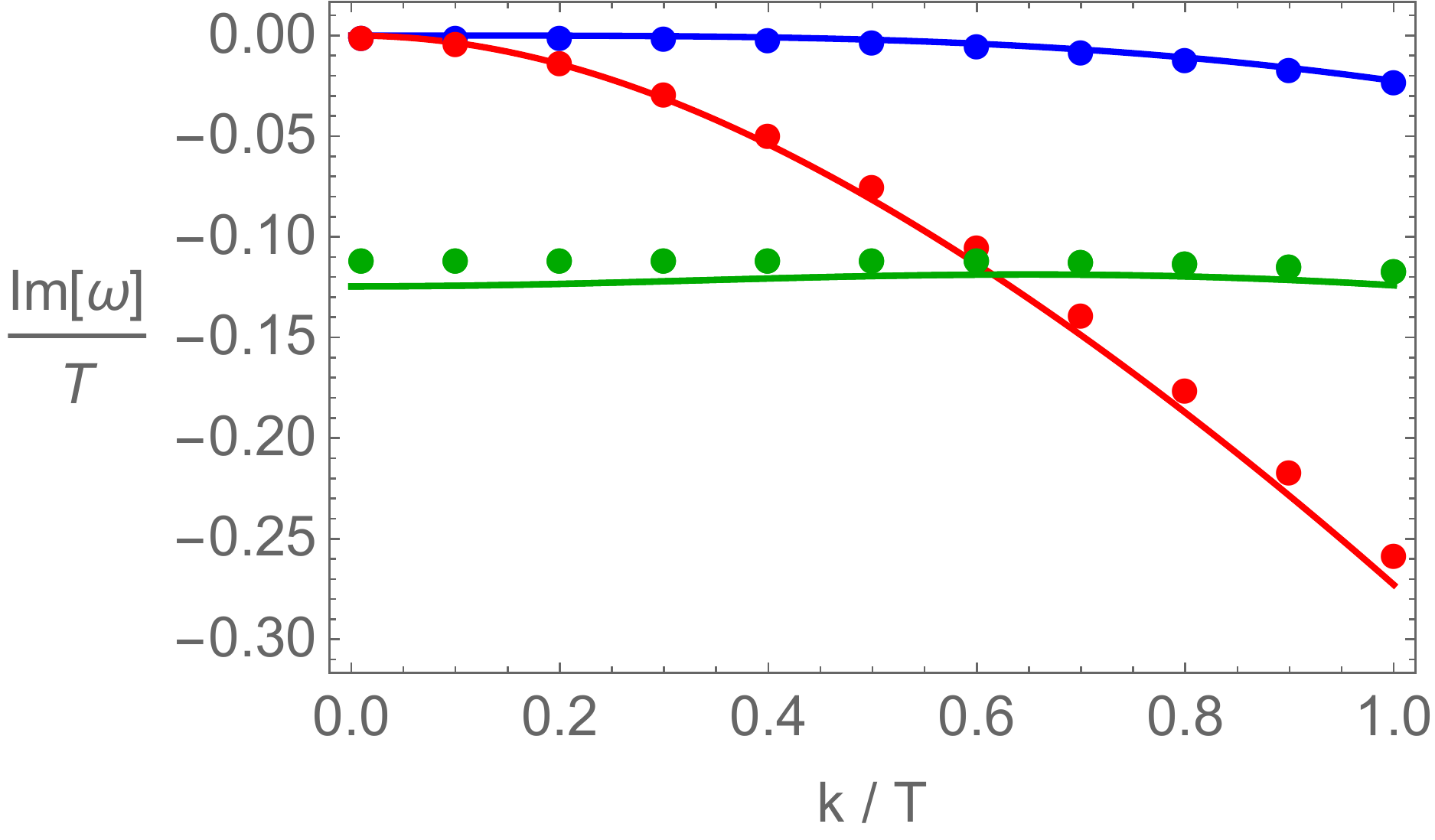} \label{QNMSFIG2f}}
 \caption{Quasi-normal modes vs dispersion relations from hydrodynamics at \textit{finite} density ($\mu/T=5$). \textbf{Left:} (a) and (d) at $H/T^2=0$. \textbf{Center:} (b) and (e) at $H/T^2=5$. \textbf{Right:} (c) and (f) at $H/T^2=10$. \textbf{All figures:} colored dots are numerically computed quasi-normal modes and solid lines are dispersion relations from hydrodynamics by \eqref{DETDISPER}.  }\label{QNMSFIG2}
\end{figure}

\paragraph{Quasi-normal modes at finite density:} Next, let us discuss the case at finite density. We display the representative quasi-normal mode data at $\mu/T=5$ at Fig. \ref{QNMSFIG2} and compare them with dispersion relations from hydrodynamics: see also Table. \ref{ST2}.
\begin{figure}[]
\centering
     {\includegraphics[width=7.2cm]{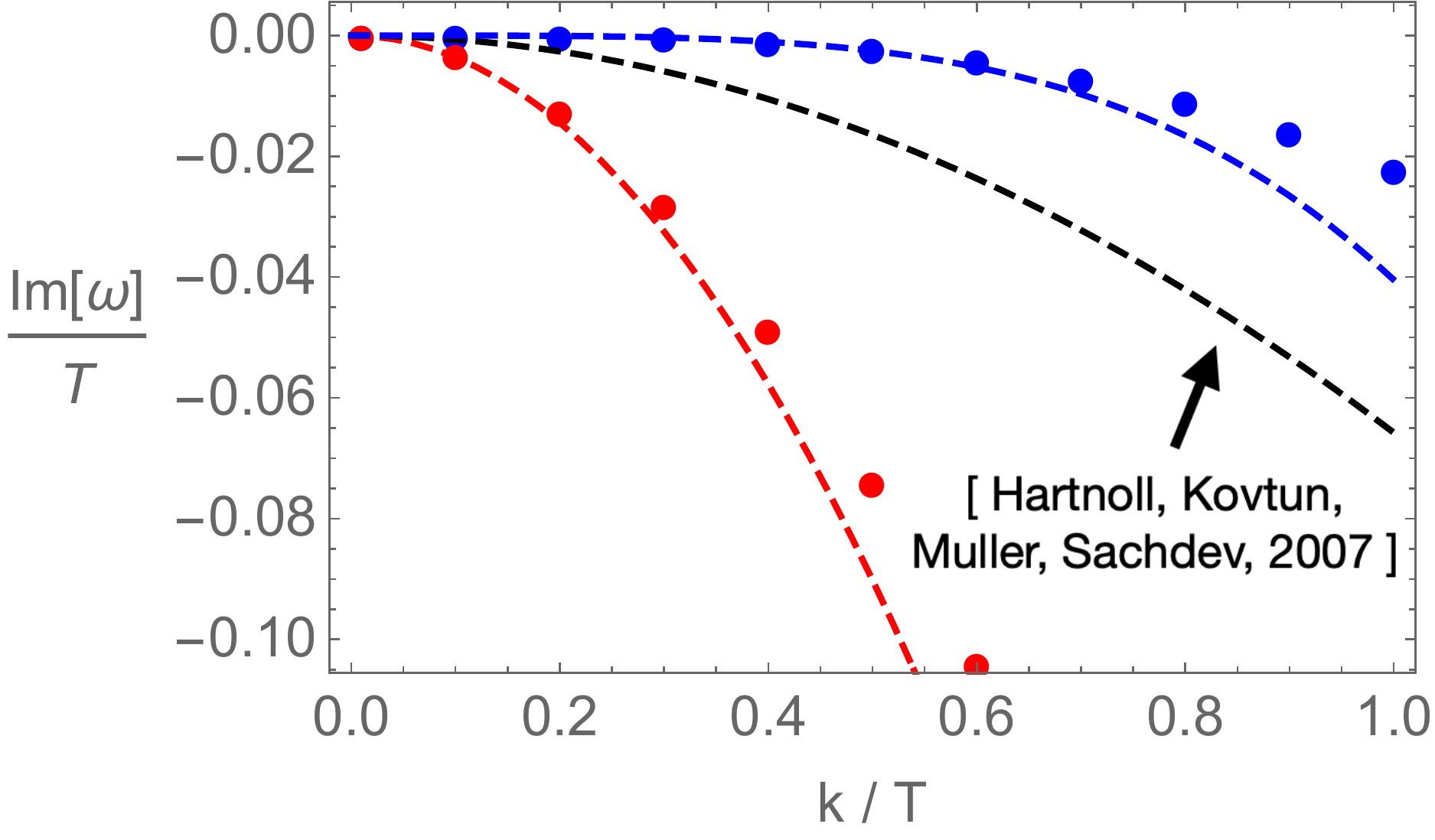} \label{}}
 \caption{Gapless modes at $\mu/T=5$, $H/T^2=10$: Fig. \ref{QNMSFIG2f}. All dots are quasi-normal modes. Red dashed line is the diffusion mode \eqref{disfin1} and the blue dashed line is sub-diffusion mode \eqref{disfin2}. Black dashed line is the diffusion mode given in \cite{Hartnoll:2007ih}.}\label{QNMSFIG3}
\end{figure}
For $H=0$, (a) and (d), the green and red data correspond to \eqref{SCDSP1def} and \eqref{MAGNETO1def}, respectively. The blue data is the shear diffusion mode \eqref{MAGNETO2def}. 
For $H\neq0$, (b) and (e), (or (c) and (f)) have the diffusion mode \eqref{disfin1} (red data), the sub-diffusion mode \eqref{disfin2} (blue data), and the cyclotron mode \eqref{disfin2} (green data). Note that, as in the zero density case, quasi-normal modes are well approximated with hydrodynamics at small magnetic fields. Note also that the cyclotron mode at finite density has a real gap as well as an imaginary gap. 

As we demonstrated in the section \ref{SECMHDDIS}, the prefactor $\rhomu$ of the diffusion mode \eqref{disfin1} in its numerator was not shown in \cite{Hartnoll:2007ih}. Thus, it will be instructive to compare \eqref{disfin1} with the one given in \cite{Hartnoll:2007ih}. See Fig. \ref{QNMSFIG3}. One can find that the prefactor is important to match quasi-normal modes with hydrodynamics.

%%%%%%%%%%%%%%%%%%%%%%%%%%%%%%%%%%%%%%
%    Section: 
%%%%%%%%%%%%%%%%%%%%%%%%%%%%%%%%%%%%%%
\subsection{Diffusion bounds at finite density}\label{}

We close this section with the investigation of the transport properties of the gapless modes: the diffusion constant from the diffusion mode \eqref{disfin1} and the sub-diffusion constant from the sub-diffusion mode \eqref{disfin2}.\footnote{For the transport properties of the gapped mode, i.e., the cyclotron mode in \eqref{disfin2}, see \cite{Hartnoll:2007ih}.}
In particular, we focus on the bound of the diffusion constants. It was proposed \cite{Blake:2016wvh,Blake:2016sud} that the diffusion constant $D$ may have the lower bound as
\begin{align}\label{qcplb1}
\begin{split}
D \geq v_{B}^2/\lambda_{L}  \,,
\end{split}
\end{align}
which is associated with the properties from quantum chaos~\cite{Shenker:2013pqa,Blake:2016wvh,Roberts:2014isa,Roberts:2016wdl}:
\begin{align}\label{qcplb}
\begin{split}
v_{B}^2 = \frac{\pi T}{r_{h}}\,, \quad \lambda_{L} = 2\pi T \,,
\end{split}
\end{align}
where $v_{B}$ is the butterfly velocity and $\lambda_{L}$ is the Lyapunov exponent. 
The proposal \eqref{qcplb1} has been checked in many models~\cite{Lucas:2018wsc,Davison:2018ofp,Gu:2017njx,Ling:2017jik,Gu:2017ohj,Baggioli:2016pia,Blake:2016jnn,Blake:2017qgd,Wu:2017mdl,Li:2019bgc,Ge:2017fix,Li:2017nxh,Ahn:2017kvc,Baggioli:2017ojd,Kim:2017dgz,Baggioli:2020ljz,Jeong:2021zhz}.

\begin{figure}[]
\centering
     \subfigure[$D$ is from \eqref{disfin1}]
     {\includegraphics[width=6.43cm]{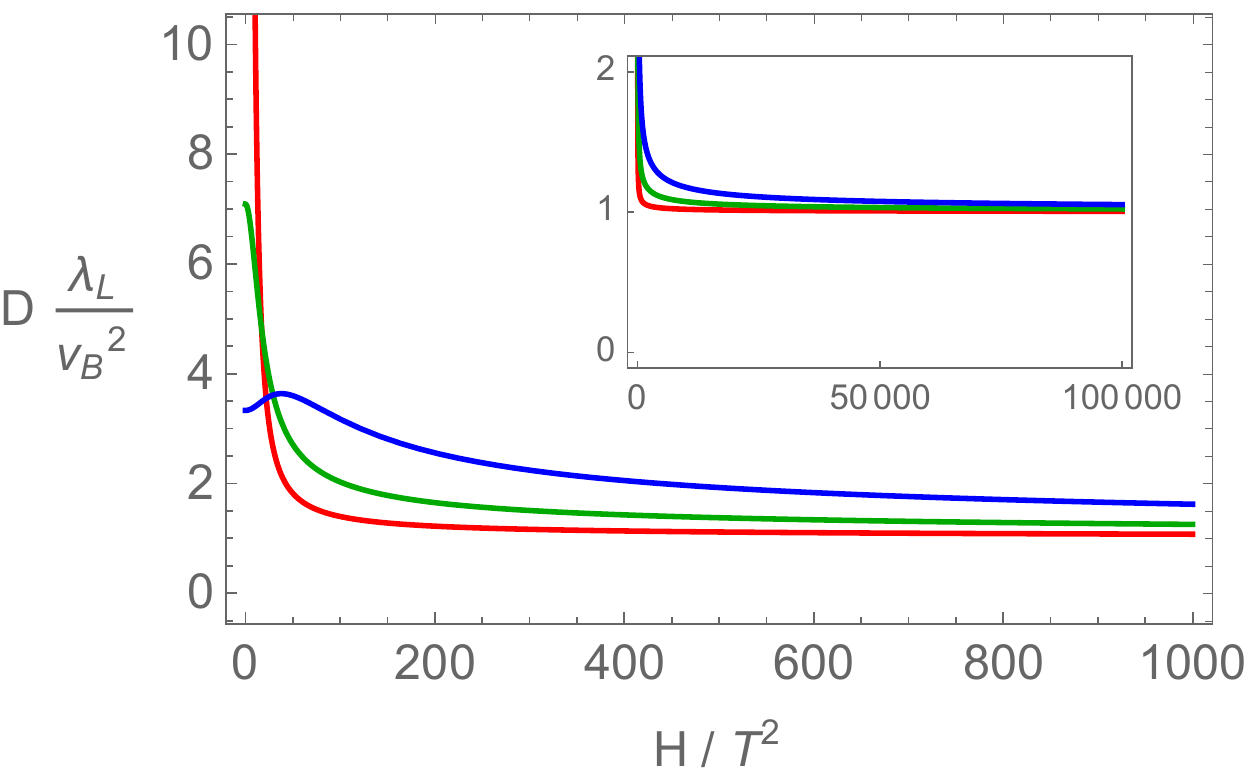} \label{DFBFIGdda}}
     \subfigure[$D_{\text{shear}}$ is from \eqref{disfin2}]
     {\includegraphics[width=6.93cm]{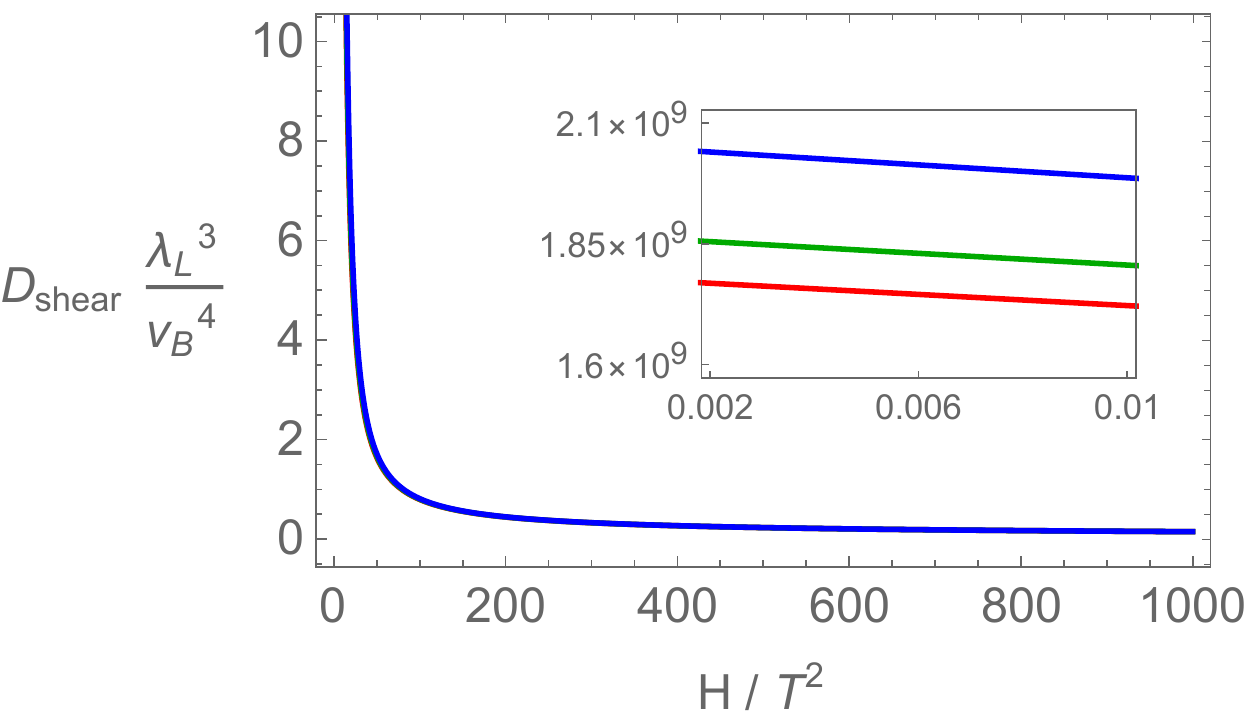} \label{}}
 \caption{Diffusion bounds at $\mu/T=(0, 3, 5)$ (red, green, blue). The diffusion constant $D$ can have the lower bound \eqref{qcplb1} at $H/T^2\gg1$, while the sub-diffusion constant $D_{\text{shear}}$ does not.}\label{DFBFIGdd}
\end{figure}

In Fig. \ref{DFBFIGdd}, we found that the diffusion constant from \eqref{disfin1} can respect the lower bound \eqref{qcplb1} in the presence of both a density and a magnetic field, while the sub-diffusion constant $D_{\text{shear}}:=\frac{\eta}{H^2 \left( \frac{\partial \rho}{\partial \mu} \right)_{T}}$ in \eqref{disfin2} may not.  
Note that the neutral case (red data) in Fig. \ref{DFBFIGdda}, is the reproduction for the result in \cite{Jeong:2021zhz}.

\paragraph{Further comments on the diffusion constant:}
We make two further comments on the diffusion constant $D$ in \eqref{disfin1}.
First, it is instructive to check if $D$ is related to the energy diffusion constant $D_{E}$ at finite density, since $D$ at \textit{zero} density was found to be the same as $D_{E}$~\cite{Jeong:2021zhz}.
The energy diffusion constant for the dyonic black hole was given \cite{Blake:2017qgd,Blake:2015hxa} as follows
\begin{align}\label{EDCdtd}
\begin{split}
D_{E} = \frac{\kappa}{c_{\rho}} \,,\quad \kappa = \frac{s^2 T H^2}{H^4 + \rho^2 H^2} \,, \quad c_{\rho} = T \left(\frac{\partial s}{\partial T}\right)_{\rho} \,, 
\end{split}
\end{align}
where thermodynamic quantities are \eqref{HAWKINGT}.

\begin{figure}[]
\centering
     \includegraphics[width=7.2cm]{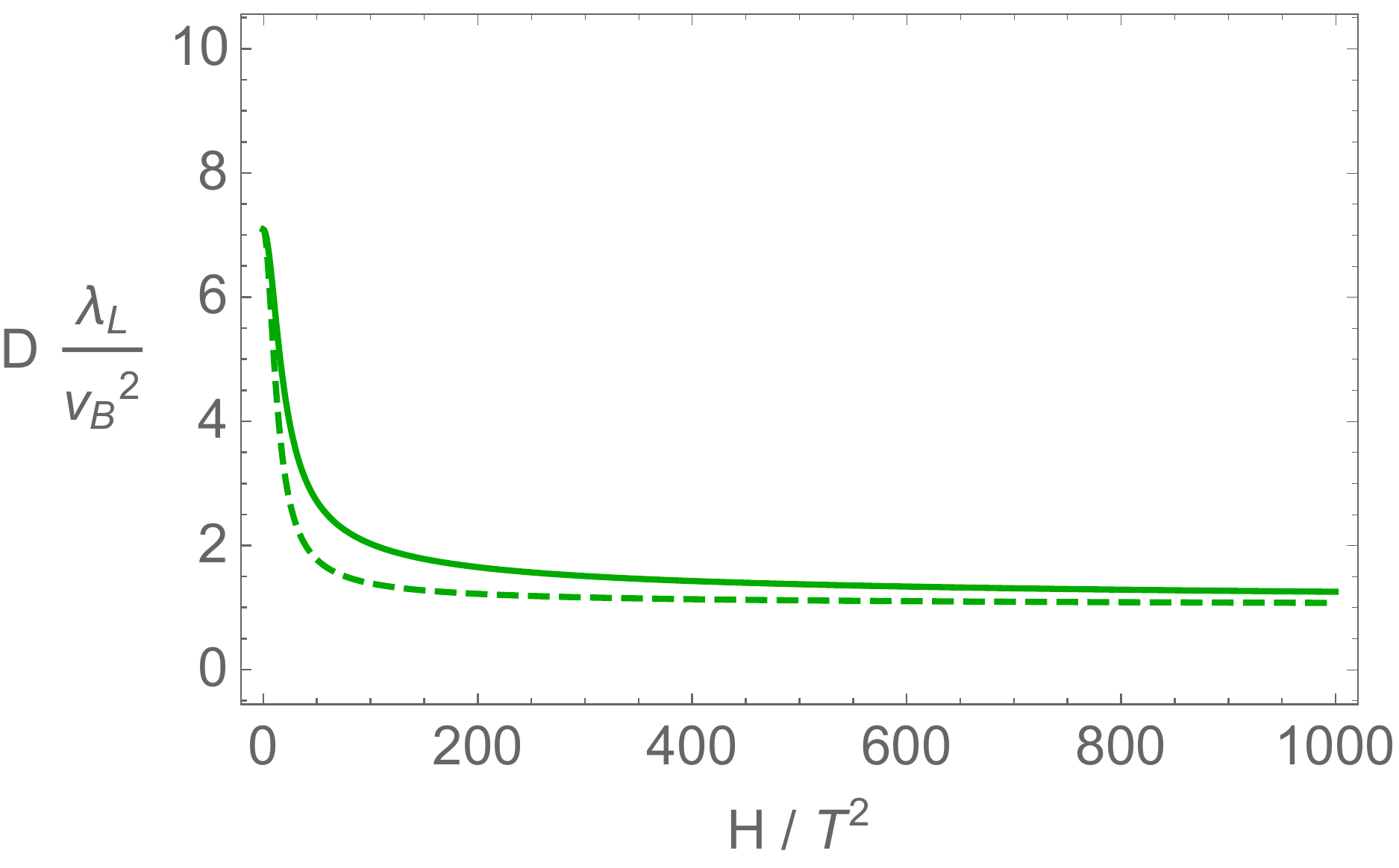} 
 \caption{Diffusion bounds at $\mu/T= 3$. Solid line is from \eqref{disfin1} and dashed line is from \eqref{EDCdtd}. }\label{DCFCOMFIG}
\end{figure}

In Fig. \ref{DCFCOMFIG}, we display both the diffusion constant in \eqref{disfin1} and the energy diffusion constant \eqref{EDCdtd} at finite density.
One can see that the diffusion (a solid line), \eqref{disfin1}, is different from the energy diffusion (a  dashed line), \eqref{EDCdtd}, in general at \textit{finite} density. 

Note that $D_{E}$ could be finite even at vanishing magnetic field (i.e., translational invariance is not broken) when the system has a density, since $\kappa$ in \eqref{EDCdtd} at $H\rightarrow0$ is finite if $\rho\neq0$. Note also that $D\sim D_{E}$ at $H/T^2\gg1$ may be consistent with \cite{Blake:2017qgd} stating that the diffusion process is governed by the energy diffusion in the low temperature limit of finite density fixed points.\footnote{It may also be consistent with axion models \cite{Kim:2017dgz} in which the diffusion constant at finite density can be identified with the energy diffusion constant in the incoherent regime ($m/T\gg1$). Here $m$ is the axion charge describing the strength of momentum relaxation.}

Second, from the recent development of quantum chaos, it was also suggested \cite{Blake:2018leo} that the lower bound of the diffusion constant, \eqref{qcplb1}, may be associated with the phenomena from the ill-defined Green's function, called pole-skipping~\cite{Grozdanov:2017ajz,Blake:2017ris,Blake:2018leo}. 
In particular, pole-skipping states that there is a special point in the momentum space as
\begin{align}\label{PSP}
\begin{split}
\omega = \omega_{*} \,, \quad k = k_{*}\,,
\end{split}
\end{align}
in which the Green's function $G(\omega_{*}, k_{*})\sim\frac{0}{0}$, i.e., ill-defined or not uniquely determined.\footnote{See \cite{Blake:2019otz,Grozdanov:2019uhi,Ceplak:2019ymw,Ceplak:2021efc,Natsuume:2019sfp,Natsuume:2019xcy,Natsuume:2019vcv,Grozdanov:2018kkt,Grozdanov:2020koi,Li:2019bgc,Liu:2020yaf,Abbasi:2020ykq,Jansen:2020hfd,Wu:2019esr,Abbasi:2019rhy,Haehl:2018izb,Das:2019tga,Ramirez:2020qer,Ahn:2019rnq,Ahn:2020bks,Ahn:2020baf,Choi:2020tdj,Kim:2020url,Sil:2020jhr,Abbasi:2020xli,Jeong:2021zhz,Kim:2021hqy,Kim:2021xdz,Blake:2021hjj,Mahish:2022xjz} for the recent development of pole-skipping.}

In \cite{Jeong:2021zhz}, it has been found that the leading pole-skipping point \eqref{PSP} of the gravitational sound channel for the generic holographic model including the dyonic black holes \eqref{ACTIONH} is 
\begin{align}\label{PSP2}
\begin{split}
\omega_{*} = i\,\lambda_{L} \,, \quad k_{*} = i\,\frac{\lambda_{L}}{v_{B}} \,,
\end{split}
\end{align}
where the quantum choas properties are \eqref{qcplb}.
With \eqref{PSP2}, the lower bound of the diffusion constant \eqref{qcplb1} was realized by that the hydrodynamic diffusion mode, $\omega=-i D k^2$, e.g. \eqref{disfin1}, is passing through the pole-skipping point \eqref{PSP2} at low temperature as 
\begin{align}\label{PSP3}
\begin{split}
\omega_{*} = -i\,D\,k_{*}^{2} \quad\rightarrow\quad D = v_{B}^2/\lambda_{L} \,.
\end{split}
\end{align}

To the best of our knowledge, the pole-skipping argument \eqref{PSP3} for the lower bound of the diffusion constant only has been confirmed at \textit{zero} density cases in literature: the energy diffusion with the axion model~\cite{Blake:2018leo} or with a magnetic field~\cite{Jeong:2021zhz}, the crystal diffusion~\cite{Jeong:2021zhz}.

Thus, in order to develop the proposal \eqref{PSP3} further, it will be important to check if such an argument holds even at \textit{finite} density. 
For this purpose, we investigate if the lower bound of the diffusion constant found in Fig. \ref{DFBFIGdd}  can be related to the pole-skipping \eqref{PSP2}.
\begin{figure}[]
\centering
     \subfigure[$H/T^2=10$]
     {\includegraphics[width=6.73cm]{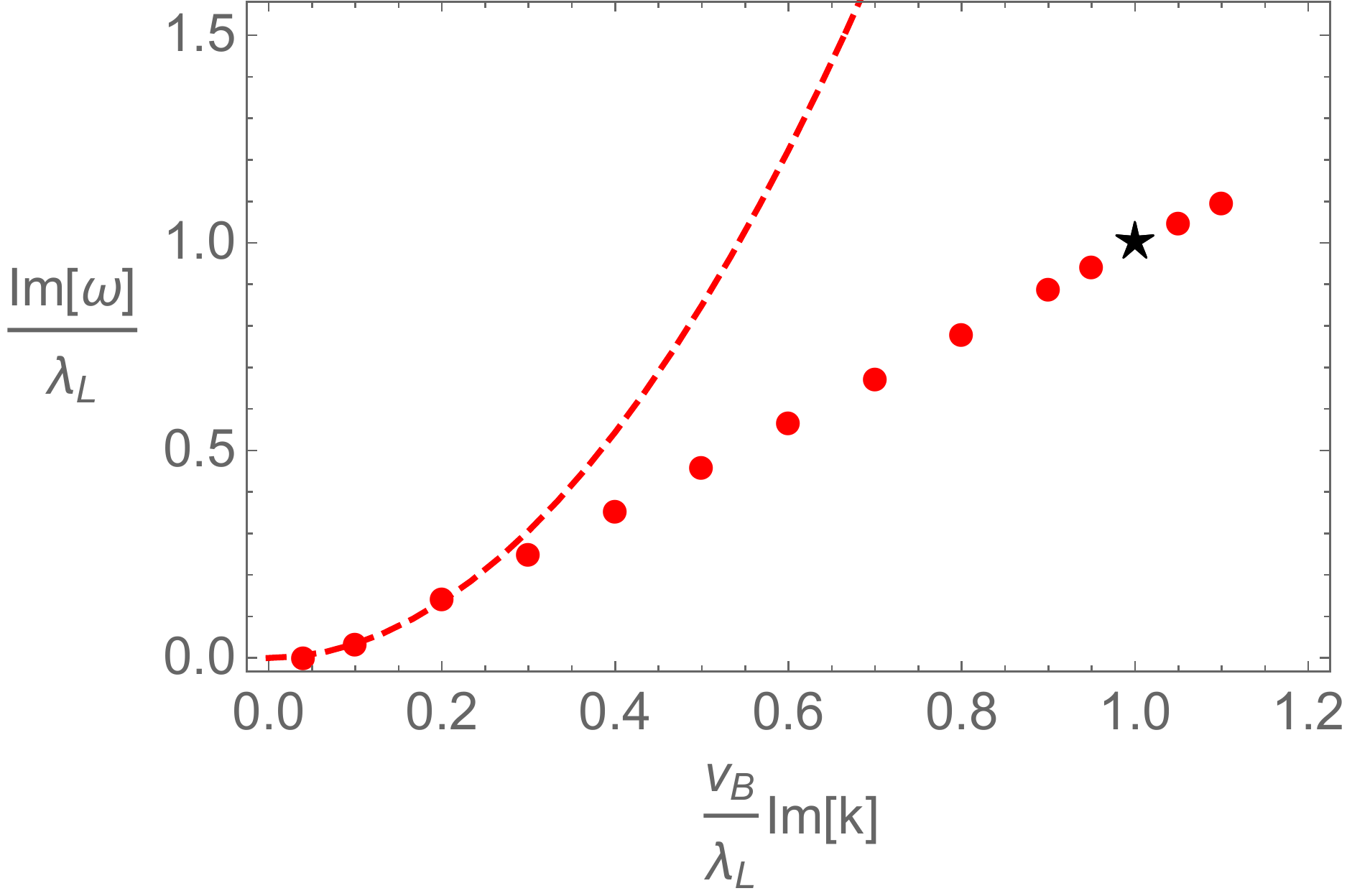} \label{PSPFIG1}}
     \subfigure[$H/T^2=10000$]
     {\includegraphics[width=6.73cm]{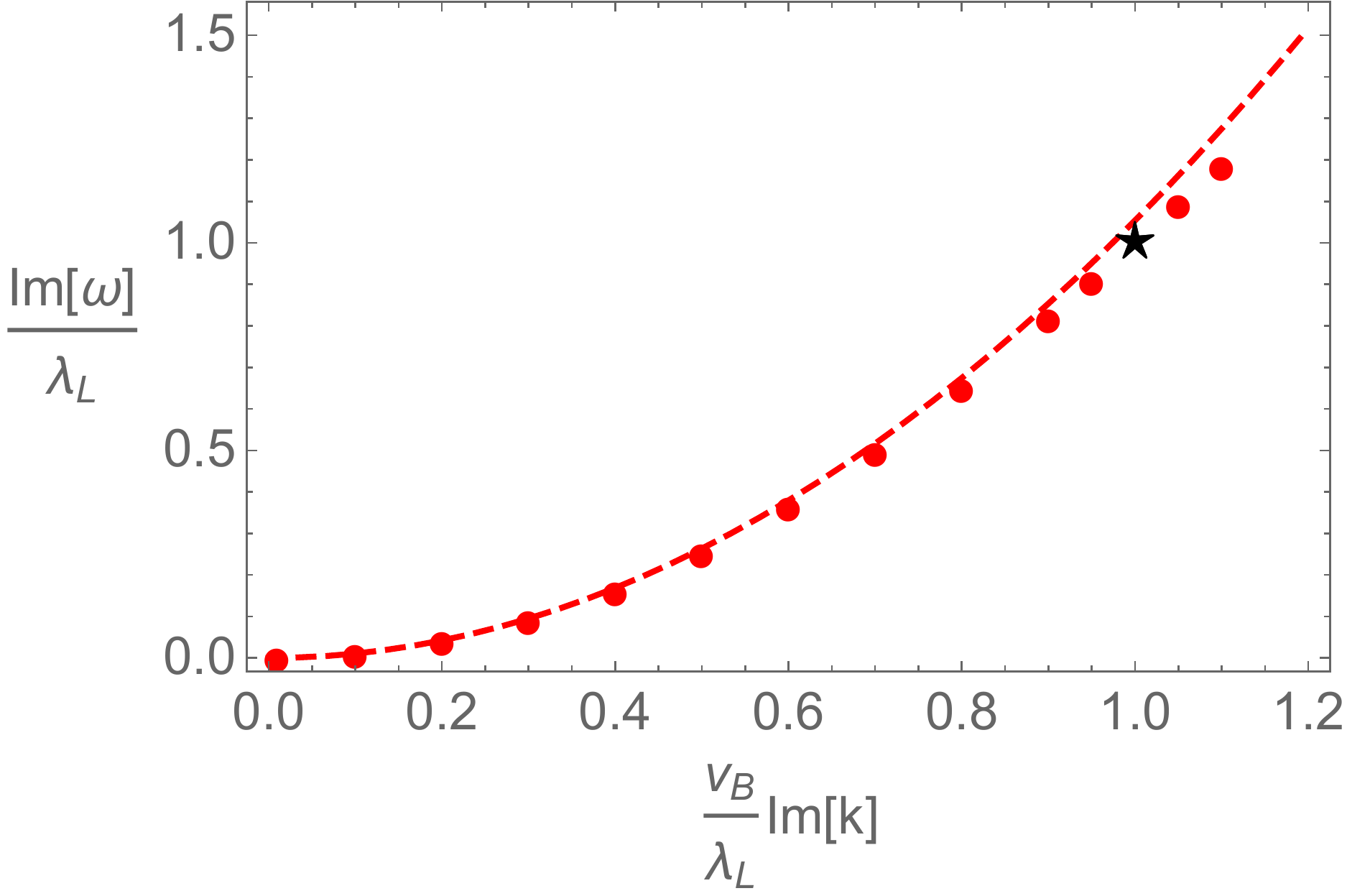} \label{PSPFIG2}}
 \caption{Pole-skipping and diffusive mode at $\mu/T=5$. All dots are quasi-normal mode, a dashed line is \eqref{disfin1}, and the star is the pole-skipping point \eqref{PSP2}.}\label{PSPFIG}
\end{figure}

In Fig. \ref{PSPFIG}, we find that the pole-skipping argument \eqref{PSP3} holds even at finite density: as $H/T^2\gg1$ (low temperature limit) from Fig. \ref{PSPFIG1} to Fig. \ref{PSPFIG2}, the pole-skipping point \eqref{PSP2} is passing through the diffusive mode \eqref{disfin1}. One may consider Fig. \ref{PSPFIG} to be a direct generalization of \cite{Jeong:2021zhz} to the case of a finite density.

%%%%%%%%%%%%%%%%%%%%%%%%%%%%%%%%
%   Conclusion 
%%%%%%%%%%%%%%%%%%%%%%%%%%%%%%%%
\section{Conclusion}\label{SECMHDDIS444}

We have studied the quasi-normal modes of the dyonic black holes in (3+1) dimensions.
In particular, we also revisited the Hartnoll-Kovtun-M\"uller-Sachdev (HKMS) magneto-hydrodynamics in (2+1) dimensions~\cite{Hartnoll:2007ih} and checked that the quasi-normal modes of dyonic black holes are consistent with the dispersion relations from HKMS magneto-hydrodynamics.

Furthermore, from the detailed analysis of the HKMS magneto-hydrodynamics we slightly corrected the dispersion relation given in previous literature~\cite{Hartnoll:2007ih}, which is important for the matching with quasi-normal modes.
Within the quasi-normal mode computations in holography, we also found the relevant independent fluctuation variables \eqref{GIVOUR} of the dyonic black holes, which was not present in previous literature.
For the summary of the dispersions of dyonic black holes, see Table. \ref{ST1} (the neutral case) and Table. \ref{ST2} (finite density case).

Our work not only provides another successful example showing the consistency between quasi-normal modes in (3+1) dimensions and the hydrodynamic predictions in (2+1) dimensions along the line of \cite{Davison:2014lua,Amoretti:2017frz,Andrade:2017cnc,Amoretti:2018tzw,Ammon:2019wci,Amoretti:2019cef,Amoretti:2019kuf,Baggioli:2021xuv,Jeong:2021zhz,Amado:2009ts,Herzog:2009md,Yarom:2009uq,Herzog:2011ec,Amado:2013xya,Amado:2013aea,Esposito:2016ria,Arean:2021tks,Ammon:2021slb} in holography, but also is useful for the complete understanding of the dyonic black holes in that our work extends the previous works, the thermodynamic properties or the transport properties at \textit{zero} wave vector, of the dyonic black holes~\cite{Hartnoll:2007ih,Hartnoll:2007ip,Hartnoll:2007ai,Herzog:2007ij,Denef:2009yy,OBannon:2007in,Buchbinder:2008dc,Buchbinder:2009aa,Buchbinder:2009mk,Bergman:2012na,Gubankova:2013lca,Dutta:2013dca,Kim:2015wba,Amoretti:2015gna,Blake:2015ina,Lucas:2015pxa,Zhou:2015dha,Davison:2015bea,Blake:2015hxa,Donos:2015bxe,Seo:2015pug,Ahn:2015shg,Amoretti:2016cad,Kim:2016hzi,Ge:2016sel,Khimphun:2017mqb,Blake:2017qgd,Cremonini:2017qwq,Seo:2017yux,Chen:2017gsl,Blauvelt:2017koq,Angelinos:2018qlc,Pal:2019bfw,Kim:2019lxb,Hoyos:2019pyz,Song:2019rnf,Amoretti:2019buu,Baggioli:2020edn,An:2020tkn,Kim:2020ozm22,Amoretti:2020mkp,Amoretti:2021lll,Amoretti:2021fch,Jokela:2021uws,Priyadarshinee:2021rch} for the case at \textit{finite} wave vector.

In addition to matching quasi-normal modes with the hydrodynamic theory, we also investigated the transport property at finite wave vector: the diffusion constant. We found that the diffusion constant from the dyonic black hole can have a lower bound at low temperature and show that such a lower bound can also be understood as the pole-skipping. In particular, our work confirmed the relation between the diffusion bound and pole-skipping at a \textit{finite} density for the first time.

One of the interesting future directions from this paper will be to investigate the \textit{dynamical} gauge fields of dyonic black holes. In particular, following \cite{Hernandez:2017mch} considering the (3+1) dimensional hydrodynamics of the dynamical gauge fields, one can also study the (2+1) dimensional hydrodynamics together with the dynamical gauge field and compare it with the quasi-normal modes of dyonic black holes~\cite{wipYW3}.\footnote{One may realize the dynamical gauge field in holography, for instance by the alternative quantization. More details will be given in \cite{wipYW3}.}

It may also be interesting to study the quasi-normal modes of the dyonic black holes in the presence of the explicitly broken translational invariance. For instance, the dyonic black holes with the axion model~\cite{Kim:2015wba} produce the DC conductivities (i.e., zero wave vector property) 
\begin{align}\label{OINT1}
\begin{split}
(m=0):\quad& \sigma^{xx}_{DC}=\sigma^{yy}_{DC}=0 \,, \qquad\qquad\,\,\,\, \sigma^{xy}_{DC}=-\sigma^{yx}_{DC} = \frac{\rho}{H} \,, \\
(H=0):\quad& \sigma^{xx}_{DC}=\sigma^{yy}_{DC}=1+\frac{\mu^2}{m^2} \,, \qquad \sigma^{xy}_{DC}=-\sigma^{yx}_{DC} = 0 \,, \\
\end{split}
\end{align} 
where $m$ is the strength of the translational symmetry breaking, $H$ is the magnetic field. 
One can see that the two limits given in \eqref{OINT1} do not commute. This implies that magneto-hydrodynamics with the broken translational symmetry may also give different dispersion relations (i.e., the finite wave vector property) depending on whether we take $m=0$ first or $H=0$ first. Thus, the interplay between HKMS magneto-hydrodynamics, the first line in \eqref{OINT1}, and  hydrodynamics with  broken translational invariance, the second line in \eqref{OINT1}, may not be a trivial subject. Note that one can also study  similar topics with  spontaneously broken symmetry~\cite{Baggioli:2020edn}.
We leave these subject as future work and hope to address it in the near future.

%%%%%%%%%%%%%%%%%%%%%%%%%%%%%%%%
%    Section: Acknowledgments
%%%%%%%%%%%%%%%%%%%%%%%%%%%%%%%%
\acknowledgments

We would like to thank  {Yongjun Ahn, Matteo Baggioli, Kyoung-Bum Huh}  for valuable discussions and correspondence.  
This work was supported by the National Key R$\&$D Program of China (Grant No. 2018FYA0305800), Project 12035016 supported by National Natural Science Foundation of China, the Strategic Priority Research Program of Chinese Academy of Sciences, Grant No. XDB28000000, Basic Science Research Program through the National Research Foundation of Korea (NRF) funded by the Ministry of Science, ICT $\&$ Future Planning (NRF- 2021R1A2C1006791) and GIST Research Institute(GRI) grant funded by the GIST in 2022.

%%%%%%%%%%%%%%%%%%%%%%%%%%%%%%%%
%    Section: END
%%%%%%%%%%%%%%%%%%%%%%%%%%%%%%%%

\bibliographystyle{JHEP}

\providecommand{\href}[2]{#2}\begingroup\raggedright\endgroup

\end{document}